\documentclass[10pt, journal]{IEEEtran}
\usepackage{times,amsmath,amssymb,amsfonts,amsthm,graphicx,epsfig,cite,psfrag}
\usepackage{algpseudocode,algorithmicx,algorithm}
\usepackage{color}
\usepackage{subcaption}
\def\gammabf{\boldsymbol \gamma }

\def\lambdabf{\boldsymbol \lambda }

\def\phibf{\boldsymbol \phi }
\def\varphibf{\boldsymbol \varphi }

\def\abf{{\bf a}}
\def\bbf{{\bf b}}
\def\cbf{{\bf c}}

\def\ebf{{\bf e}}

\def\gbf{{\bf g}}
\def\hbf{{\bf h}}

\def\qbf{{\bf q}}

\def\wbf{{\bf w}}
\def\xbf{{\bf x}}
\def\ybf{{\bf y}}

\def\xbf{{\bf x}}
\def\ybf{{\bf y}}
\def\Abf{{\bf A}}

\def\Cbf{{\bf C}}

\def\Gbf{{\bf G}}
\def\Hbf{{\bf H}}
\def\Ibf{{\bf I}}

\def\Rbf{{\bf R}}

\def\Vbf{{\bf V}}

\def\Ac{{\cal A}}

\def\Cc{{\cal C}}

\def\Ec{{\cal E}}

\def\Gc{{\cal G}}

\def\Kc{{\cal K}}

\def\Mc{{\cal M}}
\def\Nc{{\cal N}}

\def\Pc{{\cal P}}

\def\Sc{{\cal S}}

\def\Yc{{\cal Y}}

\def\eg{{\it e.g.,\ \/}}

\def\ie{{\it i.e.,\ \/}}
\def\nn{\nonumber}

\def\SINR{\textrm{SINR}}

\def\tot{\text{tot}}
\def\d{\text{d}}
\def\r{\text{r}}
\def\B{\text{B}}

\def\st {\text{s.t.}}

\newcommand{\mbbE}{\mathbb{E}}
\theoremstyle{definition}

\newtheorem{proposition}{Proposition}
\newtheorem{remark}{Remark}

\begin{document}

\title{Fast and Scalable  Beamforming  for RIS-Assisted Downlink Multi-group Multicasting}
\author{ Mohammad Ebrahimi, Min Dong, \IEEEmembership{Fellow, IEEE}, and Mitra Hekmat\thanks{The authors are with the Dept. of Electrical, Computer, and Software Engineering, Ontario Tech University, Ontario, Canada. Preliminary result of this work was presented in \cite{EbrahimiDong:Asilomar23}.}}

\maketitle

\begin{abstract}
This paper considers downlink multi-group multicasting via beamforming facilitated by a reconfigurable intelligent surface (RIS). We  develop a fast  and scalable algorithm for  the joint  base station (BS)  and  RIS beamforming optimization to  minimize the transmit power while meeting user quality-of-service (QoS) targets. By analyzing the structure of the QoS constraints, we reformulate the problem and show that the joint beamforming optimization  inherently consists of  a  multicast beamforming QoS problem for the BS and a passive multicast beamforming max-min-fair (MMF) problem for the RIS. We propose a fast alternating multicast beamforming (AMBF) algorithm to effectively solve the two subproblems alternatingly. For the BS multicast subproblem, we  utilize the optimal  multicast beamforming structure to efficiently determine the BS beamformers. For the RIS multicast subproblem,  we reformulate the MMF problem and apply a first-order projected subgradient algorithm (PSA), which yields simple closed-form updates. The computational complexity of the AMBF algorithm grows  linearly with the number of RIS elements and BS antennas. We further consider joint BS and RIS beamforming for the weighted MMF design objective subject to the BS transmit power budget.\ We propose an alternating PSA (APSA) fast algorithm to  compute
the beamforming solutions for the BS and RIS. APSA consists only  closed-form updates per iteration, yielding linear computational complexity in the number of RIS elements and BS antennas. 
Simulation results show the efficacy of our proposed algorithms in terms of   performance and computational cost compared to alternative methods.  \end{abstract}


\maketitle

\section{Introduction} \label{sec:Introduction}
\IEEEPARstart{R}{econfigurable}   intelligent surface (RIS) is an emerging next-generation technology that can  actively control and enhance wireless propagation channel conditions, thereby creating a smart  reconfigurable wireless environment to enhance  communications performance \cite{Liu&etal:COMSvTut21,Wuetal:TCOM21}. RIS uses a planar surface consisting of passive reflective elements to control the phase shifts of the reflected wireless signal towards desired directions with minimal energy consumption. This forms passive beamforming, which enhances the transmission performance. RIS offers several advantages over the traditional methods (\eg base station (BS) or relay), such as the elimination of the need for complex signal processing,  low energy consumption, reduced hardware cost, and convenience for deployment, etc. The potential role of RIS in next-generation wireless  networks has been investigated to support  a variety of applications, including  coverage extension, capacity enhancement, and improved localization and sensing \cite{Bjorson:SPM22,Wu&etal:WCOM19,Huang&etal:WCOM19, Guo&etal:WCOM20,Tishchenko:ComTut25}.

Existing and emerging wireless services and technologies, such as video streaming, software dispatch distribution,   edge storage and computing, and distributed machine learning, are driving significant growth of downlink  content and data distribution. This demand increases the need of wireless multicast strategies to support efficiency delivery of shared content. In particular, with a multi-antenna BS, physical-layer multicasting via beamforming is an efficient transmission technique  for the BS to deliver common data  to multiple users simultaneously, which can effectively improve spectrum and power efficiency.
Numerous optimization algorithms or signal processing techniques have been  developed \cite{Sidiropoulosetal:TSP06,Karipidisetal:TSP08,Ottersten&etal:TSP14,TranHanifJuntti:SPL14,KonarSidiropoulos:TSP17,Chen&Tao:COM17,Sadeghietal&Sanguinetti:TWC17,Yu&Dong:ICASSP18,Yu&Dong:SPAWC18,
 Mohamadietal:TSP22,Mohamadietal:TSP24,Dong&Wang:TSP2020,Zhang&etal:COML22,
 Zhang&Dong&Liang:TSP23,Mohammadi&Dong&SS:TSP21,YinDong:TCOM25}. For delivering common data to a group, multicast beamforming performance
is typically limited by the user in the group with the worst channel
condition. This limitation arises from the wireless prorogation environment. RIS can be utilized  to enhance    multicast transmissions by adding an additional controllable channel path to mitigate this condition. By adjusting the phase shifts of the reflective elements, RIS enables passive beamforming that enhances the channel
conditions for users, thereby   improving the overall transmission performance.
Therefore, it is crucial to explore effective and efficient designs for RIS-assisted multicast beamforming.

Downlink multicast beamforming presents a challenging design problem, as it is generally a non-convex NP-hard problem \cite{Sidiropoulosetal:TSP06}. Earlier literature  adopted the semi-definite relaxation (SDR) approach \cite{Sidiropoulosetal:TSP06,Karipidisetal:TSP08,Ottersten&etal:TSP14} to find approximate solutions. However, with the increasing number of BS antennas in evolved wireless systems, the successive convex approximation (SCA) technique has gained popularity due to its improved performance and lower computational complexity \cite{TranHanifJuntti:SPL14}. To further reduce the solution complexity in massive multiple-input multiple-output (MIMO) systems, various computational methods or signal processing techniques have been developed for single-cell scenarios \cite{KonarSidiropoulos:TSP17,Chen&Tao:COM17,Sadeghietal&Sanguinetti:TWC17}, multi-cell cooperation \cite{Yu&Dong:ICASSP18,Yu&Dong:SPAWC18}, as well as robust designs that account for imperfect channel state information \cite{Mohamadietal:TSP22,Mohamadietal:TSP24}.  The optimal structure for multi-group multicast beamforming  has been established recently in \cite{Dong&Wang:TSP2020}. It indicates that to design the BS beamformers, we only need to optimize the weights among users in each group. This significantly simplifies the optimization problem, as it is much smaller and independent of the number of BS antennas.
This optimal structure has been leveraged to develop scalable ultra-low-complexity   algorithms for  massive MIMO downlink multicasting   \cite{Zhang&etal:COML22,Zhang&Dong&Liang:TSP23},  mixed traffic scenarios \cite{Mohammadi&Dong&SS:TSP21}, and multi-cell coordination\cite{YinDong:TCOM25}.

The challenge of multicast beamforming design extends to RIS-assisted downlink multicasting. RIS-assisted multicast beamforming design has  been investigated for the single-group  \cite{Du&etal:WCOM21,Tao&etal:WCL21} and multi-group settings \cite{Zhou&etal:TSP20,Farooq&etal:Globecom22,Li&etal:VTC20,Shu&etal:VTC21}. Several approaches have been
proposed for designing joint BS and RIS beamforming \cite{Zhou&etal:TSP20,Farooq&etal:Globecom22,Li&etal:VTC20,Shu&etal:VTC21}. The alternating optimization (AO) technique is commonly adopted for  joint BS and RIS beamforming optimization. 
Joint BS and RIS beamforming optimization for maximizing the sum group rate is considered in \cite{Zhou&etal:TSP20,Farooq&etal:Globecom22}. In  \cite{Zhou&etal:TSP20}, a  smoothing technique is proposed along with a majorization minimization method to solve the subproblems in each AO\ iteration.
In \cite{Farooq&etal:Globecom22}, the smoothing technique  is applied directly to the rate objective, along with a low-complexity alternating projected gradient
algorithm.
However, while maximizing  the sum group rate can enhance overall data rate, it does not guarantee fairness among different groups. As a result, some groups may experience significant performance degradation, leading to an  unacceptable user experience. To tackle the QoS issue, the QoS problem that minimizes the BS transmit power while meeting user QoS targets  is studied in \cite{Li&etal:VTC20},  where  AO is directly applied to the  optimization problem, resulting in a feasibility subproblem for the RIS passive beamforming design. The SDR-based approach is then proposed to solve the non-convex subproblems in each AO iteration.
A similar approach is also considered in  \cite{Shu&etal:VTC21} for an RIS-assisted symbiotic radio multicast system comprising  the primary receivers and an internet-of-things receiver. However, solving the feasibility problem does not ensure the quality of the solution for the RIS passive beamforming subproblem, raising questions about the effectiveness of this direct AO approach.
Additionally, the methods mentioned above generally entail high computational complexity, which hinders  their practical implementation. Most of these studies have only considered a relatively small number of RIS elements for performance evaluation. Since RIS systems typically require hundreds of elements to be beneficial, it is crucial to develop a scalable design solution that balances both high performance and low computational complexity.

\subsection{Contribution}
 This paper aims to provide an effective joint BS and RIS beamforming design for RIS-assisted downlink data multicasting.  We develop fast and scalable algorithmic solutions for two multicast beamforming formulations: the QoS problem, which minimizes the BS transmit power and the max-min fair (MMF) problem, which maximizes the weighted minimum  signal-to-interference-and-noise ratio (SINR)  among users, subject to the BS transmit power limit. 


We propose a fast alternating multicast beamforming (AMBF) approach  for solving the QoS problem. This approach differs from existing methods that directly apply the AO technique to the joint optimization problem. Instead, we explore the structure of  the constraints    and reformulate the
problem to reveal that 
the QoS problem inherently consists of two
separate multicast beamforming optimization subproblems: a conventional  multicast beamforming QoS problem for the BS and a  passive multicast beamforming MMF problem for the RIS.
The proposed AMBF approach solves these two subproblems alternatingly,  effectively improving the BS and RIS beamforming solutions over  iterations. 

We design fast algorithms to effectively solve the two subproblems.  For the BS multicast problem, we  utilize the optimal  multicast beamforming structure \cite{Dong&Wang:TSP2020} to  compute a solution efficiently, where the computational complexity is independent of the number of BS antennas. This approach contrasts with existing methods for RIS-assisted beamforming, which rely on computational optimization techniques.  
For the more challenging RIS passive multicast MMF problem,
we  reformulate the problem to employ a  first-order projected subgradient algorithm (PSA), which yields computationally inexpensive closed-form updates and  has a guaranteed convergence.
In the special case of the multi-user unicast beamforming scenario, the optimal BS beamformer can be computed exactly, and  our AMBF approach requires only closed-form solutions or updates for both subproblems. Our AMBF algorithm is scalable, with computational complexity increasing linearly with the number of RIS elements and BS antennas.

For the MMF problem, we show that it is an inverse problem to the earlier QoS problem. We leverage the optimal BS beamforming structure to transform the MMF problem into an equivalent lower-dimensional problem and propose an alternating PSA (APSA) algorithm that efficiently computes the BS and RIS beamforming solutions.  Our APSA algorithm employs closed-form updates for both BS and RIS beamformers, maintaining linear computational complexity with respect to (w.r.t.) the number of RIS elements and BS antennas.

Simulation results show the effectiveness of our proposed algorithms,  demonstrating superior  performance  compared to  existing AO-based algorithms and comparative to SCA-based joint optimization methods, while achieving significant reduction in    computational time.

\subsection{Organization and Notations}
The rest of this paper is organized as follows. Section \ref{sec:model} presents the system model and  the problem formulation for the RIS-assisted multicast beamforming QoS problem. In Section \ref{sec:Prop_Alg}, we develop our AMBF approach to break  the QoS problem into two subproblems to solve alternatingly. In Section \ref{sec:AMBF-alg}, we present the fast algorithms to solve each beamforming subproblem.   The MMF problem is considered in Section \ref{sec:MMF}, where we discuss the relation between the QoS and MMF problems and propose the  APSA  algorithm to solve the MMF problem efficiently. The simulation
results is provided in Section~\ref{sec:sim}, followed by the conclusion
in Section~\ref{sec:conclusion}. 

\textit{Notations:} The symbols used to represent Hermitian, transpose, and conjugate are $(.)^H$, $(.)^T$, and $(.)^*$, respectively. The Euclidean norm of a vector is symbolized by $\|.\|$.  The abbreviation i.i.d. stands for independent and identically distributed, and   $\xbf\sim \Cc\Nc(0,\Ibf)$ represents a complex Gaussian random vector with zero mean and covariance $\Ibf$.

\allowdisplaybreaks
\section{System Model } \label{sec:model}
\subsection{System Model}
\begin{figure}[t]
\centering
\psfrag{RIS}[c][r]{\small RIS: $M$ elements}
\psfrag{controller}[c][c]{\small RIS controller}
\psfrag{antenna}[c][r]{\small : $N$ antennas}
\psfrag{Hr}[c][r]{\small $\Hbf_\r$}
\psfrag{hr1}[c][l]{\small $\{\hbf^\r_{1k}\}$}
\psfrag{hr2}[c][l]{\small $\{\hbf^\r_{2k}\}$}
\psfrag{hd1}[c][b]{\small $\{\hbf^\d_{1k}\}$}
\psfrag{hd2}[c][c]{\small $\{\hbf^\d_{2k}\}$}
\psfrag{w}[l][l]{\small $\{\wbf_1,\wbf_2\}$}
\psfrag{e}[l][l]{\small $\ebf$}
\includegraphics[width= .8\columnwidth]{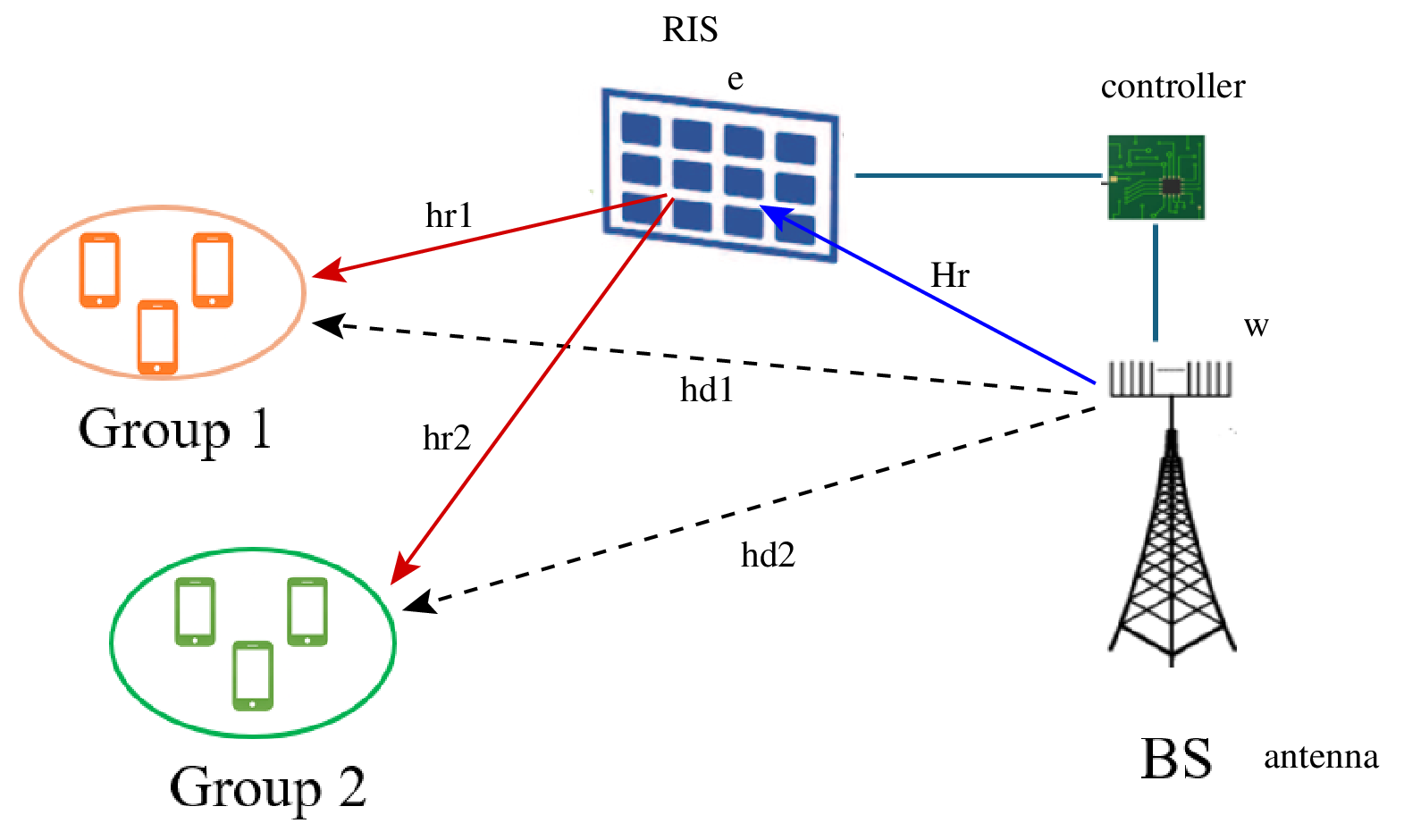}
\caption{An example of RIS-assisted downlink multi-group multicasting. }
\label{fig:sys_model}
\end{figure}

We consider an  RIS-assisted downlink multicast scenario: a BS  equipped with $N$ antennas multicasts messages to $G$ user groups, and an RIS consisting of  $M$ passive reflective elements is deployed to assist the data transmission between the BS and the users, as shown in Fig.~\ref{fig:sys_model}.
   We assume that  group $i$ consists of $K_i$ single-antenna users, and the BS sends a common message to this group of users that  is independent of other groups. The total number of users is denoted by $K_{\tot}=\sum_{i=1}^GK_i$. The BS controls the phases of the RIS array elements via a RIS controller to adjust the  signals received at the RIS to the desired directions.
We denote the set of   group indices by  $\Gc \triangleq \{1, \ldots, G\}$,  the set of user indices in group $i$ by  $\Kc_i \triangleq \{1, \ldots, K_i\}$, $i\in \Gc$, and the set of RIS element indices by $\Mc \triangleq\{1,\ldots,M\}$.  

 Our design considers a general scenario where both the direct path and the RIS path are available between the BS and the users,\footnote{In a rich scattering environment, although there may not be a line-of-sight path available, the non-line-of-sight paths typically still exist  between the BS and a user.  } which includes the special case that only the RIS path is available.   Let  $\Hbf_\text{r}$ denote the $M \times N$ channel matrix from the BS to the RIS,  $\hbf^{\text{d}}_{ik}$  the $N\times 1$ channel vector from the BS to user $k$ in group $i$, and $\hbf^\text{r}_{ik}$  the $M \times 1$ channel vector from the RIS to user $k$ in group $i$, for $k \in \Kc, i\in\Gc$. We use  $\wbf_i$ to  represent the $N\times 1$ multicast beamforming vector at the BS for group $i \in \Gc$. Also,    $\ebf\triangleq [e_1,\ldots,e_M]^T$ represents the  vector containing the $M$ reflection coefficients at the RIS, where $e_m=e^{j\theta_m}$ with $\theta_m \in(-\pi,\pi]$ being the phase shift of element $m$. We refer to $\ebf$ as the RIS passive beamforming vector. The signal received at  user $k$ in group $i$ is given by
\begin{align*}
\hspace*{-.5em}y_{ik} 
 &=\! \big(\hbf^\text{d}_{ik}+\Hbf_\text{r}^H\text{diag}(\ebf)\hbf^\text{r}_{ik}\big) ^{\!H}\!\sum_{j=1}^G\wbf_js_j \!+n_{ik} \nn\\
&= \!\big(\hbf^\text{d}_{ik}\!+\! \Gbf_{ik}\ebf\big)^{\!H}\wbf_i s_i +\! \sum_{j\neq i}\!\big(\hbf^\text{d}_{ik}\!+\! \Gbf_{ik}\ebf\big)^{\!H}\wbf_j s_j \!+\! n_{ik}
\end{align*}
where $s_i$ is the  symbol intended for group $i$ with $\mbbE[|s_i|^2] = 1$,  $\Gbf_{ik}  \triangleq \Hbf_\text{r}^H\text{diag}(\hbf_{ik}^{\text{r}})$ is the $N\times M$ cascaded channel from the BS to user $k$ in group $i$ via the RIS, $n_{ik}\sim\Cc\Nc(  0,\sigma^2)$ is the receiver additive white Gaussian noise with variance $\sigma^2$.

The received SINR at user $k$ in group $i$ is given by 
\begin{align}\label{eqn:SINR}
\SINR_{ik} = \frac{|\big(\hbf^\text{d}_{ik}+ \Gbf_{ik}\ebf\big)^H\wbf_i |^2}{\sum_{j\neq i}|\big(\hbf^\text{d}_{ik}+ \Gbf_{ik}\ebf\big)^H\wbf_j |^2+\sigma^2},
\end{align}
and the total transmit power at the BS is  $ \sum_{i=1}^G \|\wbf_i\|^2$. 

\subsection{Problem Formulation}
We first focus on the QoS problem for the  RIS-assisted multicast beamforming design. Our objective is to jointly  optimize the multicast beamforming vectors $\wbf_i$ and RIS reflection coefficient vector $\ebf$ to minimize the BS transmit power while ensuring that the SINR target at each user is met. This joint  optimization problem is formulated as
\begin{subequations}
\begin{align}
\Pc_o: \  \min_{\wbf, \ebf}  & \ \ \sum_{i=1}^G \|\wbf_i\|^2 \label{Obj_Power}\\
\st & \ \ \SINR_{ik} \ge \gamma_{ik}, \ k\in \Kc_i, ~i \in \Gc \label{SINR_constr}\\
& \ \ |e_m|^2=1 , \ m \in \Mc  \label{Unit_Mod_constr}
\end{align}
\end{subequations}
where $\wbf\triangleq[\wbf_1^H,\ldots,\wbf_G^H]^H$, and $\gamma_{ik}$ is the SINR target for user $k$ in group $i$. Note that the SINR constraint in~\eqref{SINR_constr}  is equivalent to  the per-user minimum rate constraint;  it  essentially defines a  QoS target for each user. When $\gamma_{ik}=\tilde{\gamma}_i$, $\forall k$,   this target becomes the group rate target, and constraint~\eqref{SINR_constr}  reduce to ensure that  the  minimum group rate target $\tilde{\gamma}_i$ is met for group $i$.

Problem $\Pc_o$ is a non-convex quadratically constrained quadratic programming (QCQP)  problem. For  conventional downlink transmission without  RIS, the multicast beamforming problem w.r.t. $\wbf$ is  known to be NP-hard. Therefore, the joint optimization problem $\Pc_o$ is even more challenging to solve, since it includes the additional RIS passive beamformer $\ebf$  affecting   $\SINR_{ik}$  to be optimized under the non-convex unit-modulus constraint \eqref{Unit_Mod_constr}.  Furthermore,  the number of RIS elements $M$ is expected to be large to realize RIS benefits. Thus,   $\Pc_o$ is typically a large-scale problem to solve. Our goal is to develop a fast and scalable algorithmic solution that is of high performance for RIS-assisted multicast beamforming. 

\section{ RIS-assisted Multicast Beamforming Design}\label{sec:Prop_Alg}
A common approach to solve $\Pc_o$ is to consider using alternating optimization (AO)  to compute the BS multicast beamformer $\wbf$ and RIS passive beamformer $\ebf$  alternatingly. However, directly applying AO  to  $\Pc_o$ is not an effective method. This is because  the objective is only a function of  $\wbf$. Given  $\wbf$,  $\Pc_o$ is reduced to a feasibility problem w.r.t. $\ebf$, which cannot ensure the quality of solution.  For example, the value of $\ebf$ from the previous AO iteration is already  a feasible solution. The algorithm could stuck at the value of $(\wbf,\ebf)$  after the first round of updates, and the quality of this solution  merely depends on the quality of the initial point. Thus, instead of directly applying AO, we  first analyze the structure of  $\Pc_o$ w.r.t. $\wbf$ and $\ebf$ and then utilize it to devise an effective algorithm to solve   $\Pc_o$.
\subsection{Reformulation }\label{subsec:QoS}
 To explore the structural relation between $\wbf$ and $\ebf$ in  $\Pc_o$, we first convert the set of SINR\ constraints in \eqref{SINR_constr} into the following equivalent constraint:
\begin{align}
\SINR_{ik} \ge \gamma_{ik}, \  \forall k, \forall i  
\quad &  \Leftrightarrow  \quad \frac{\SINR_{ik}}{\gamma_{ik}} \ge 1, \  \forall k, \forall i  \nn \\ 
&\Leftrightarrow \quad \min_{i,k} \frac{\SINR_{ik}}{\gamma_{ik}} \ge 1. \label{SINR_constr:mod}
\end{align}
Replacing \eqref{SINR_constr} with \eqref{SINR_constr:mod}, we can equivalently rewrite $\Pc_o$ as
\begin{subequations}
\begin{align}
\Pc_1: \  \min_{\wbf, \ebf}  & \ \ \sum_{i=1}^G \|\wbf_i\|^2 \\
\st & \ \ \min_{i,k} \frac{\SINR_{ik}}{\gamma_{ik}} \ge 1 \label{SINR_constr:mod1}\\
& \ \ |e_m|^2=1 , \ m \in \Mc. \label{Unit_mod_constr1}
\end{align}
\end{subequations}

Noticing that  RIS passive beamforming vector $\ebf$ only affects $\SINR_{ik}$'s in  \eqref{SINR_constr:mod1}, we can combine \eqref{SINR_constr:mod1} and \eqref{Unit_mod_constr1} and further  transform $\Pc_1$ into the following equivalent problem:
\begin{subequations}\label{P2}
\begin{align}
\Pc_2: &\  \min_{\wbf}   \ \ \sum_{i=1}^G \|\wbf_i\|^2  \\
\st & \ \ \max_{\substack{\ebf: |e_m|^2=1 \\  m \in \Mc}}\min_{i,k} \frac{\SINR_{ik}}{\gamma_{ik}} \ge 1. \label{SINR_constr:mod2}
\end{align}
\end{subequations}
To see the equivalence, it is straightforward to show  $\Pc_1$ and $\Pc_2$ have the same feasible set. 

From $\Pc_o$ to $\Pc_2$, the original set of SINR constraints in  \eqref{SINR_constr} is now replaced by the constraint  \eqref{SINR_constr:mod2}. Note that the left hand side of this new constraint is a max-min weighted SINRs  optimization problem w.r.t. $\ebf$, and the unit-modulus constraints on $\ebf$ 
in \eqref{Unit_Mod_constr} are now the constraints for this max-min optimization problem. Thus, the joint optimization problem $\Pc_o$ is transformed into $\Pc_2$, which inherently contains two separate optimization subproblems w.r.t. $\wbf$ and $\ebf$, respectively. Following this, we  can naturally break down $\Pc_2$ into   two subproblems to solve alternatingly.
\subsection{Proposed Alternating  Multicast Beamforming Approach } \label{AMBA}
Examining the two subproblems in  $\Pc_2$, we will show below that they are essentially two types of multicast beamforming problems:  1) BS multicast beamforming QoS problem w.r.t. $\wbf$; 2) RIS passive multicast beamforming MMF problem w.r.t. $\ebf$. We solve these two subproblems  alternatingly.
Our proposed alternating multicast beamforming (AMBF) approach includes two steps: 1) the alternating optimization step, and 2) a final processing step.  The final processing step is to ensure a feasible solution to $\Pc_o$. 
Below, we  will  focus on  describing the two subproblems and the final processing step, leaving the algorithm design for solving the subproblems to Sections~\ref{subsec:P_e} and \ref{subsec:S_w}.

\subsubsection{BS Multicast Beamforming QoS Problem for $\wbf$}\label{subsec:w_problem}
Given RIS passive beamforming vector $\ebf$, $\Pc_o$ is reduced to the following  BS multicast beamforming problem w.r.t. $\wbf$:
\begin{subequations} \label{Pe}
\begin{align}
\label{ObjW_Power}
\Pc_{\ebf}: \  \min_{\wbf}  & \ \ \sum_{i=1}^G \|\wbf_i\|^2 \\
\label{SINRW_constr}
\st & \ \ \SINR_{ik} \ge \gamma_{ik}, \ k\in \Kc_i, ~i \in \Gc,
\end{align}
\end{subequations}
which is   a conventional  downlink multi-group multicast beamforming QoS problem. The effective channel between the BS and  user $k$ in group $i$ consists of the direct path and the RIS-path given by  $\hbf^\text{d}_{ik}+ \Gbf_{ik}\ebf$, as shown in $\SINR_{ik}$ in \eqref{eqn:SINR}.  

\subsubsection{RIS Passive Multicast Beamforming MMF Problem for $\ebf$}\label{subsec:RIS_problem}
Given $\wbf$,  RIS passive beamformer $\ebf$  optimization is the main challenge in $\Pc_o$. However, once we convert $\Pc_o$ into $\Pc_2$, 
we see that optimizing  $\ebf$ is essentially to maximize the minimum ratio of the received SINR at each user over its target, \ie the weighted SINR, as the following max-min problem:
\begin{subequations}
\begin{align}
\label{v_MMF}
\Sc_{\wbf}: &\ \ \max_{\ebf}\min_{i, k}
 \ \ \frac{\SINR_{ik}}{\gamma_{ik}}\\
\label{Unit_Mod_v}
\st & \ \  |e_m|^2=1 , \ m \in \Mc 
\end{align}
\end{subequations}
where $\SINR_{ik}$ in \eqref{eqn:SINR} can be re-expressed as
\begin{align}\label{eqn:SINR_e}
\SINR_{ik} = \frac{|\ebf^H\Gbf_{ik}^H\wbf_i + \hbf^\text{d}_{ik}\!^H\wbf_i |^2}{\sum_{j\neq i}| \ebf^H\Gbf_{ik}^H\wbf_j + \hbf^\text{d}_{ik}\!^H\wbf_j|^2+\sigma^2}.
\end{align}
\begin{remark}\label{remark1}
Note that $\Sc_{\wbf}$  can be viewed as a variant of single-group multicast beamforming weighted MMF    problem, with $\ebf$ being the multicast beamformer  to all the users.  This can be more clearly seen  in the RIS-path only case, given by
\begin{align}\label{eqn:SINR_e_2}
\SINR_{ik} = \frac{|\ebf^H\Gbf_{ik}^H\wbf_i |^2}{\sum_{j\neq i}| \ebf^H\Gbf_{ik}^H\wbf_j|^2+\sigma^2}.
\end{align}
 But it has a few differences  from the conventional MMF multicast beamforming: i) It contains  self-interference, since     $\ebf$ appears in the denominator of   $\SINR_{ik}$ in \eqref{eqn:SINR_e_2}. 
{ii)} From the perspective of  RIS beamformer  $\ebf$,  $\SINR_{ik}$ in \eqref{eqn:SINR_e_2} can  be interpreted as the signal-to-leakage-and-noise ratio (SLNR) for user $k$ in group $i$. The power leakage is through the  effective channels formed by other BS beamformers $\wbf_j$'s to the same user, $\Gbf_{ik}^H\wbf_j$, $\forall j\neq i$. Thus,  the MMF problem $\Sc_{\wbf}$ w.r.t. $\ebf$ is based on an  SLNR metric. For the general case with a direct path   in \eqref{eqn:SINR_e}, the combined beamed signal strength from the RIS and the direct path need to be considered. {iii)} Different from the conventional total transmit power constraint, the constraints in \eqref{Unit_Mod_v} can be viewed as a non-convex per-element  power constraint, where each element is required to consume exactly one unit of power.  
\end{remark}

Due to the difference discussed above, $\Sc_\wbf$ is a more challenging problem to solve than the conventional MMF problem, which is a non-convex QCQP problem. To tackle this problem, we  
note that the constraints in \eqref{Unit_Mod_v} are equivalent to 
\begin{align} \label{Unit_v:eq}
|e_m|^2 \le 1, m \in \Mc  \quad\text{and}\quad \ebf^T\ebf=M.
\end{align}
Thus, we  replace the constraints in \eqref{Unit_Mod_v} with  \eqref{Unit_v:eq} and transform $\Sc_\wbf$ into the following equivalent problem:
\begin{subequations}
\begin{align} 
\Sc_{\wbf}^\text{eq}: \ \max_{\ebf}\min_{i, k}
& \ \ \frac{\SINR_{ik}}{\gamma_{ik}} \label{Unit_Mod_v_mod}\\
\st & \ \  |e_m|^2 \le 1, \ m \in \Mc \label{Unit_Mod_v_a}\\
& \ \ \ebf^T\ebf=M. \label{Unit_Mod_v_b}
\end{align}
\end{subequations}

 To make the problem more tractable, we  relax $\Sc_\wbf^\text{eq}$ by   transferring the constraint in   \eqref{Unit_Mod_v_b} into the objective function in \eqref{Unit_Mod_v_mod} as a penalty term with a penalty weight $\rho>0$. The relaxed problem is given by
\begin{subequations}\label{v_MMF:0}
\begin{align}\label{v_MMF:1}
\widetilde{\Sc}_{\wbf}: \ \max_{\ebf}& \ \bigg(\!\min_{i, k}
  \frac{\SINR_{ik}}{\gamma_{ik}}\bigg) + \rho\frac{\ebf^H\ebf}{M}\\
\st & \ \  |e_m|^2 \le 1, \ m \in \Mc. \label{Relax_Unit_Mod_v}
\end{align}
\end{subequations} 
\begin{remark} \label{remark2}
The penalty term  in \eqref{v_MMF:1} is to ensure that  $\ebf^T\ebf$ is as close to $M$ as possible in solving $\widetilde{\Sc}_{\wbf}$.  Note that we choose the penalty term  in this particular form such that  the  two terms in the objective function  are both normalized. In particular, for  first term, we note that $\wbf$ is the solution to $\Pc_\ebf$, which yields  $\min_{i, k}\frac{\SINR_{ik}}{\gamma_{ik}} =1$ in \eqref{SINRW_constr}. Thus, we expect  this term to be close to $1$, especially when the iterative procedure between $\Pc_\ebf$ and $\widetilde{\Sc}_\wbf$  converges. The   second term for the normalized RIS  passive beamformer  satisfies  $\frac{\ebf^H\ebf}{M}\le 1$ based on \eqref{Relax_Unit_Mod_v}, and it is also expected to be close to $1$ when the AO procedure converges.
Therefore, the two terms in \eqref{v_MMF:1} have comparable values  that are both close to $1$. Being normalized, they  do not depend on  the actual values of SINR or $\|\ebf\|^2$ for a specific system configuration. This simplifies the tuning of penalty parameter $\rho$, which can remain the same for different system configurations.
\end{remark}

Compared with $\Sc_{\wbf}$,  $\widetilde{\Sc}_{\wbf}$  is  more amenable to efficient algorithm design to compute a solution, which will be discussed in Section~\ref{subsec:S_w}. 

Combining Sections~\ref{subsec:w_problem} and \ref{subsec:RIS_problem}, the first step in our AMBF approach is to  solve the two multicast beamforming problems alternatingly, \ie  $\Pc_\ebf$ for the BS beamformers $\{\wbf_i\}$ and  $\widetilde{\Sc}_{\wbf}$ for the RIS beamformer $\ebf$.

\subsubsection{Final Processing}  Let  $(\wbf^\star, \ebf^\star)$ denote the solution from the above AO procedure. Since $\widetilde{\Sc}_{\wbf}$ is a relaxed problem of $\Sc_{\wbf}$, $(\wbf^\star, \ebf^\star)$ may not be feasible to $\Pc_o$. Thus, based on $(\wbf^\star, \ebf^\star)$, we have the final processing step to obtain the feasible solution $(\wbf^{\text{final}}, \ebf^{\text{final}})$ as follows:
\begin{itemize}
\item Project $\ebf^\star$ onto its feasible set: $\ebf^{\text{final}} = \exp(j\angle \ebf^\star)$, which is  taking the phase component of each element in $\ebf^\star$. 
\item Solve $\Pc_{\ebf^\text{final}}$ with  $\ebf^{\text{final}}$ and obtain $\wbf^{\text{final}}$.
\end{itemize}

\section{Fast  Algorithms for  alternating multicast beamforming }\label{sec:AMBF-alg}
We now propose our fast algorithm for  each of the multicast subproblems $\Pc_\ebf$ and $\widetilde{\Sc}_{\wbf}$ under the proposed AMBF approach.
\subsection{ BS  Multicast Beamforming  QoS Problem   $\Pc_\ebf$}\label{subsec:P_e}

  As mentioned earlier, $\Pc_\ebf$ is a conventional downlink multicast beamforming  QoS problem. Define $\tilde{\hbf}_{ik} \triangleq (\hbf^\text{d}_{ik}+ \Gbf_{ik}\ebf)$ as the $N \times 1$ equivalent channel  from BS to  user $k$ in group $i$ that includes both the direct path and the RIS path. 
The received SINR in \eqref{eqn:SINR} at user $k$ in group $i$ can be rewritten as 
\begin{align}\label{eqn:SINR expression}
\SINR_{ik} = \frac{|\tilde{\hbf}_{ik}^H\wbf_i|^2}{\sum_{j\neq i}^{G}|\tilde{\hbf}_{ik}^H\wbf_j|^2+\sigma^2}, \ k\in\Kc_i, i\in\Gc.
\end{align}

Despite that  $\Pc_\ebf$ is  an NP-hard non-convex QCQP problem, the  structure of the optimal   solution is obtained  in  \cite{Dong&Wang:TSP2020}, which can be utilized to compute the beamforming solution $\wbf$ with high computational efficiency, especially when the number of BS antennas is large $N \gg 1$. We directly employ this optimal structure to compute  $\wbf$. Specifically, the optimal $\wbf_i$ is a weighted MMSE  beamforming given by \cite{Dong&Wang:TSP2020}
\begin{align}
\label{W_Opt}
\wbf_i \, = \, \Rbf^{-1}(\lambdabf)\widetilde{\Hbf}_i\abf_i \, , \; i \in \Gc
\end{align}
where $\widetilde{\Hbf}_i \triangleq [\tilde{\hbf}_{i1}, \ldots,\tilde{\hbf}_{iK}]$ is the equivalent channel matrix for users in group $i$,  $\abf_{i}$ is the $K \times 1$ optimal  weight vector for group $i$ containing the (complex) weight for each user channel in the group, and  $\Rbf(\lambdabf) \triangleq {\bf{I}}+\sum_{i=1}^{G}\sum_{k=1}^{K}\lambda_{ik}\gamma_{ik}\tilde{\hbf}_{ik}\tilde{\hbf}_{ik}$  is the noise-plus-channel covariance matrix with $\lambda_{ik}$ being the optimal Lagrange multipliers associated with the SINR constraint in \eqref{SINRW_constr} and $\lambdabf$ being the vector containing all $\lambda_{ik}^o$'s.

The solution $\wbf_i$ in \eqref{W_Opt} is given in a semi-closed form with  $\lambdabf$ and $\{\abf_i\}$ to be numerically determined. We can use the algorithm proposed in  \cite{Dong&Wang:TSP2020} to compute (suboptimal) values of $\lambdabf$ efficiently using the fixed-point method. Given  $\lambdabf$ and $\wbf_i$ in \eqref{W_Opt}, $\Pc_\ebf$ can be transformed into  a  weight optimization problem w.r.t. weight vectors $\{\abf_i\}$, which is a much smaller problem with $\sum_{i=1}^GK_i$ variables compared to the original problem $\Pc_\ebf$ with $GN$ variables for $N \gg K_i$. To solve the non-convex weight optimization problem for $\{\abf_i\}$, we can adopt the fast algorithm proposed in  \cite{Zhang&Dong&Liang:TSP23}. It is  based on the SCA approach and uses an ADMM-based algorithm  to provide each SCA update using closed-form expressions with ultra-low complexity. We omit the details and refer the readers to \cite{Dong&Wang:TSP2020,Zhang&Dong&Liang:TSP23} for these algorithms.
   
\subsection{ RIS Passive Multicast Beamforming MMF Problem  $\widetilde{\Sc}_\wbf$}\label{subsec:S_w}
The  max-min problem $\widetilde{\Sc}_{\wbf}$ is expected to be large-size, since  the number of RIS elements $M$ is large. Existing methods  for solving $\ebf$  often adopt certain convexification methods, such as SDR or SCA    \cite{Zhou&etal:TSP20,Shu&etal:VTC21}. However, they suffer from  high computational complexity for large $M$. Given this subproblem needs to be solved in each AO\ iteration., it is essential to develop  an effective  low-complexity algorithm to find a solution to $\widetilde{\Sc}_{\wbf}$. 

From Remark~\ref{remark1}, we note that $\widetilde{\Sc}_{\wbf}$ is a variant of the MMF problem. We propose to apply the projected subgradient algorithm (PSA) \cite{Polyak:book1987} to  $\widetilde{\Sc}_{\wbf}$. PSA is a first-order  iterative algorithm. It  has been adopted in\cite{Zhang&etal:COML22} to solve the conventional multi-group multicast beamforming MMF problem efficiently. The structure of the objective function and constraints in     $\widetilde{\Sc}_\wbf$   are different from the conventional MMF problem. Nonetheless, we can still employ the essential technique of this method to compute a near-stationary solution of   $\widetilde{\Sc}_{\wbf}$ efficiently. Given our problem structure, applying PSA  proves to be computationally inexpensive, which is the main advantage of this approach.     

\subsubsection{Problem Reformulation}
We first transform  $\widetilde{\Sc}_\wbf$ into a min-max problem formulation. Based on the objective function in \eqref{v_MMF:1} and  using the SINR expression in \eqref{eqn:SINR}, we define 
\begin{align} \label{phi_RIS_MMF}
\hspace*{-.6em}\phi_{ik}(\ebf) \triangleq  -\bigg(\frac{1}{\gamma_{ik}} \frac{|(\hbf_{ik}^\d \!+\! \Gbf_{ik}\ebf)^H\wbf_i|^2}{\sum_{j\neq i}|(\hbf_{ik}^\d \!+\! \Gbf_{ik}\ebf)^H\wbf_j|^2\!+\!\sigma^2}  + \rho\frac{\ebf^H\ebf}{M}\bigg), 
 \end{align}
 for $k\in \Kc_i$, $i \in \Gc$. We denote the feasible set of $\widetilde{\Sc}_{\wbf}$ as  $\Ec \triangleq \{\ebf: |e_m|^2 \le 1, \ m  \in \Mc\}$.  
Then,  $\widetilde{\Sc}_{\wbf}$ can be  rewritten as a min-max problem:  \begin{align} \label{tildeS_w}
    \widetilde{\Sc}_{\wbf}:  \min_{\ebf\in\Ec} \max_{i, k} \phi_{ik}(\ebf).
 \end{align}
 
To tackle the inner maximization problem, which is an integer program, let  $\phibf(\ebf)$ be the $K_\tot\times 1$ vector containing all $\phi_{ik}(\ebf)$'s, for $k\in \Kc_i$, $i \in \Gc$. 
Also  let $\ybf$ be a  $K_\tot\times 1$ probability vector satisfying  $\ybf \succcurlyeq 0$ and $\mathbf{1}^T\ybf = 1$. Let  $f(\ebf,\ybf) \triangleq \phibf^T(\ebf)\ybf$, and and    $\Yc \triangleq \{\ybf: \ybf \succcurlyeq 0, \mathbf{1}^T\ybf = 1\}$. Then, we can further transform problem \eqref{tildeS_w} into the following equivalent problem
\begin{align}
\label{f_MMF}
\widetilde{\Sc}_{\wbf}':\min_{\ebf \in \Ec}\max_{\ybf \in\Yc}& \ f(\ebf,\ybf).
\end{align}
Both $\Ec$ and $\Yc$ are compact convex sets, and $\Yc$ is a probability simplex. Thus, an optimal solution to the inner maximization problem in $\widetilde\Sc'_{\wbf}$ is $\ybf = [0, \cdots, 1, \cdots, 0]^T$, with 1 at a position that corresponds to $\max_{i,k}\phi_{ik}(\ebf)$, which is equivalent to the inner maximization in \eqref{tildeS_w}. 

Problem $\widetilde\Sc'_{\wbf}$ is a nonconvex-concave min-max problem, since the objective function $f(\ebf,\ybf)$ is concave in $\ybf$ but nonconvex in $\ebf$. Furthermore, function $g(\ebf)\triangleq\max_{\ybf \in\Yc} f(\ebf,\ybf)$ may not be differentiable w.r.t. $\ebf$, and thus, gradient-based methods are not applicable. We note that the structure of $\widetilde{\Sc}_{\wbf}'$ is the same as the min-max problem structure considered  in \cite{Zhang&etal:COML22} (\ie the objective function is in the form of  $f(\ebf,\ybf) \triangleq \phibf^T(\ebf)\ybf,$ and the feasible set is convex). It is shown in \cite{Zhang&etal:COML22} that in this formulation,    $f(\ebf,\ybf)$ is an $L$-smooth function of $\ebf$, and $\nabla_\ebf f(\ebf,\ybf)$ is a subgradient of $g(\ebf)$. In addition, $g(\ebf)$ is $L$-weakly convex over $\Ec$. Thus, we can adopt PSA \cite{Polyak:book1987} to find a near-stationary point of $\widetilde\Sc'_{\wbf}$.

\subsubsection{ Projected Subgradient Algorithm}\label{Sw:PSA}

PSA is an iterative algorithm. Its updating procedure at iteration $j$  given by\footnote{Note that PSA in \cite{Zhang&etal:COML22} is provided using all real-valued variables. We provide the complex-valued version of this procedure, which can be shown to be equivalent.}
\begin{align}
   \ybf^{(j)} &\in \displaystyle \mathop{\arg\max}_{\ybf \in \Yc} f(\ebf^{(j)}, \ybf); \label{PSA:y}\\
  \ebf^{(j+1)} &= \Pi_{\Ec}\big(\ebf^{(j)} - \alpha\nabla_{\ebf}f(\ebf^{(j)}, \ybf^{(j)})\big) \label{PSA:e}
\end{align}
where $\alpha > 0$ is the step size, and $\Pi_{\Ec}(\ebf)$ is the projection of  $\ebf$ onto the feasible set $\Ec$. Since the constraints in $\Ec$ is for each $e_m$, $\Pi_{\Ec}(\ebf)$ performs per-element projection as
\vspace*{-.5em}
\begin{align}\label{eqn:Pi_X}
\Pi_{\Ec}(\ebf) = \begin{cases} e_m  & \mbox{if } |e_m| \leq 1 \\
   \frac{e_m}{|e_m|} & \text{o.w}
        \end{cases}, \ \text{for} \ m \in \Mc. 
\end{align}

In \eqref{PSA:y},  $\ybf^{(j)}$ can be obtained by finding an index pair $(\hat{i}, \hat{k})$ such that $\phi_{\hat{i}\hat{k}}(\ebf)=\max_{i,k}\phi_{ik}(\ebf)$, and we have $f(\ebf,\ybf^{(j)})=\phi_{\hat{i}\hat{k}}(\ebf)$. The gradient $\nabla_{\ebf}f(\ebf, \ybf^{(j)})$ in \eqref{PSA:e} has the following expression:
\begin{align}\label{PSA:f_grad1}
&\nabla_{\ebf}f(\ebf, \ybf^{(j)})  = \nabla_{\ebf}\phi_{\hat{i}\hat{k}}(\ebf) \nn \\ 
& \quad =  - \frac{\qbf_{\hat{i}\hat{i}\hat{k}}(\ebf)I_{\hat{i}\hat{k}}\!(\ebf)-\!\sum_{j \neq \hat{i}} \qbf_{j\hat{i}\hat{k}}(\ebf)S_{\hat{i}\hat{k}}(\ebf)}{\gamma_{\hat{i}\hat{k}}I_{\hat{i}\hat{k}}^2(\ebf)}  - \frac{\rho}{M}\ebf 
\end{align}
where $I_{\hat{i}\hat{k}}(\ebf) \triangleq \sum_{j \neq \hat{i}}|(\hbf_{\hat{i}\hat{k}}^\d + \Gbf_{\hat{i}\hat{k}}\ebf)^H\wbf_j|^2 + \sigma^2$, $S_{\hat{i}\hat{k}}(\ebf) \triangleq |(\hbf_{\hat{i}\hat{k}}^\d + \Gbf_{\hat{i}\hat{k}}\ebf)^{H}\wbf_{\hat{i}}|^2$, and $\qbf_{j\hat{i}\hat{k}} (\ebf)= \Gbf_{\hat{i}\hat{k}}^H\wbf_j\wbf_j^H(\hbf_{\hat{i}\hat{k}}^\d + \Gbf_{\hat{i}\hat{k}}\ebf)$.

The updates in each PSA iteration involve  closed-form evaluations as in  \eqref{PSA:y}--\eqref{PSA:f_grad1}, with only matrix and vector multiplications. Thus, PSA is computationally cheap and  is particularly suitable for solving $\widetilde{\Sc}_{\wbf}$ with a large value of $M$.  

\subsubsection{Convergence and Stopping Criterion}\label{subsec:conv}
The convergence  analysis of PSA   in \cite{Zhang&etal:COML22}  is applicable to our problem.  It shows that  the above PSA procedure is guaranteed to converge in finite time to a point at the vicinity of a stationary point of $\widetilde{\Sc}_{\wbf}$. 

Since PSA is a subgradient method, it is not guaranteed to move toward a descending direction. 
Thus, as commonly considered, we track the best solution obtained so far in $j$ iterations, \ie the one with the minimum objective value of $\widetilde{\Sc}_{\wbf}'$ in $j$ iterations: $g_{\min}^{(j)}=\min\{g^{(1)},...,g^{(j)}\}$, where $g^{(j)}=\max_{\ybf \in\Yc} f(\ebf^{(j)},\ybf)$. Note that this is  equivalent to track the maximum objective value in \eqref{v_MMF:1} obtained so far. Based on its definition, $g_{\min}^{(j)}$ is monotonically decreasing and is bounded below. Thus, we can terminate the algorithm when $g_{\min}^{(j)}$ converges.   
\subsection{Discussion on the AMBF Algorithm}
We summarize our proposed fast AMBF algorithm  for the RIS-assisted multicast QoS problem $\Pc_o$ in Algorithm~\ref{alg1}. It consists of the proposed AMBF approach in Section~\ref{AMBA} and the  fast algorithms in Sections~\ref{subsec:P_e} and \ref{subsec:S_w} for the two subproblems, respectively. A few aspects of the algorithm are discussed below:

\begin{algorithm}[t]
\caption{Alternating Multicast Beamforming Algorithm (AMBF) for RIS-Assisted QoS Problem $\Pc_o$.}\label{alg1}
\begin{algorithmic}[1]
\State  \textbf{Initialization:} Set  feasible initial point $\ebf^{0}$; Set ${\rho}$ and $\alpha$; Set $n=0$.
\Repeat
\State With $\ebf^{n}$, solve $\Pc_\ebf$ to  obtain $\wbf^{n}$ by  using \eqref{W_Opt}.
\State  With $\wbf^{n}$, set $\ebf^{(0)}=\ebf^{n}$; Solve $\widetilde{\Sc}_{\wbf}$ using PSA updates  \eqref{PSA:y} and \eqref{PSA:e} to obtain  $\ebf^{n+1}$.
\State  Set $n \gets n+1$.
\Until convergence 
\State Set  $\ebf^{\text{final}}=\Pi_{\Ec}(\ebf^{n})$. 
\State Obtain $\wbf^{\text{final}}$ by solving  $\Pc_{\ebf^\text{final}}$.
\State \textbf{return} $(\wbf^{\text{final}}, \ebf^{\text{final}})$.
\end{algorithmic}
\end{algorithm}
\subsubsection{Initialization} \label{subsubsec:init_QoS}
The AMBF algorithm  requires an initial feasible point  $\ebf^{0}$ that  satisfies  the  constraints in \eqref{Unit_Mod_constr} for the BS multicast beamforming subproblem $\Pc_\ebf$. It can be easily obtained by generating a random  phase for each element in $\ebf^{0}$. In each AO iteration $n$,   solving $\tilde{\Sc}_\wbf$ by PSA also requires an initial point $\ebf^{(0)}$ for the  update in \eqref{PSA:y}. We should set $\ebf^{(0)}=\ebf^{n}$, \ie the solution from the previous AO iteration. This  is also used  in $\Pc_\ebf$ to obtain $\wbf^{n+1}$ in the current AO\ iteration.

\subsubsection{Computational Complexity}  \label{subsec:complexity_QoS}
Algorithm~\ref{alg1} consists of solving two optimization subproblems $\Pc_{\ebf}$ and $\widetilde{\Sc}_{\wbf}$  at each iteration. The computational complexity for solving each subproblem is discussed below:

\emph{Problem $\Pc_\ebf$:}
 As mentioned in Section~\ref{subsec:P_e}, using the optimal  beamforming structure  $\wbf$ in \eqref{W_Opt},   $\Pc_{\ebf}$ is converted into the weight optimization problem w.r.t. $\{\abf_i\}$ with  $K_{\tot}(=\sum_{i=1}^GK_i)$ variables, which can be solved by SCA  \cite{Dong&Wang:TSP2020}. If each SCA subproblem is solved using a typical interior point method \cite{Boydbook} by the standard convex solver, the  computational complexity is $O(K_{\tot}^3)$ per SCA iteration, which does not grow with either $M$ or $N$. We can adopt low-complexity fast ADMM-based algorithm  \cite{Zhang&Dong&Liang:TSP23},  which only involves  closed-form updates and has the computational complexity in the order of $O(K_{\tot}^2)$. 

\emph{Problem $\Sc_{\wbf}$:} The problem is solved using  PSA  updates in \eqref{PSA:y} and \eqref{PSA:e} per iteration. Obtaining $\ybf^{(j)}$ in \eqref{PSA:y} requires computing $\SINR_{ik}$ in \eqref{eqn:SINR} for all $K_{\tot}$ users and finding the maximum $\phi_{ik}(\ebf)$ in \eqref{phi_RIS_MMF}, which need $GK_{\tot}(N+1)+K_\tot(MN+N+4)$ flops. The required computation in \eqref{PSA:e} is mainly at computing the gradient   $\nabla_{\ebf}f(\ebf, \ybf^{(j)})$  in \eqref{PSA:f_grad1}. We note that all the terms in   \eqref{PSA:f_grad1}, \ie $\qbf_{j\hat{i}\hat{k}}$,  $I_{\hat{i}\hat{k}}(\ebf)$, and $S_{\hat{i}\hat{k}}(\ebf)$ are part of the SINR\ expression that has already been computed. Thus,  calculating $\nabla_{\ebf}\phi_{\hat{i}\hat{k}}(\ebf)$ only requires  $2GM+2M$ flops. Thus, the entire updates per iteration requires $GK_{\tot}(N+1)+K_\tot(MN+N+1)+2GM+2M$ flops, with the leading complexity being $K_{\tot}N(G+M)$ flops. Since we expect  $M\gg G$ in practical scenarios, the leading computational complexity is $K_\tot NM$.  Thus, the complexity of PSA for solving $\Sc_\wbf$ grows linearly over $M$, $N$, and $K_\tot$. 

The complexity of the final per-element projection  in \eqref{eqn:Pi_X} is $O(M)$. The overall computational complexity in each AO iteration of Algorithm~\ref{alg1} is  $O(K_{\tot}^2)+\text{const} \cdot K_{\tot}MN$, which is linear in both $M$ and $N$ and quadratic in $K_\tot$.

\subsubsection{Convergence}
For  solving $\Pc_\ebf$ and $\widetilde{\Sc}_\wbf$ alternatingly, note that the objective of the minimization problem  $\Pc_\ebf$ is lower bounded and that of the maximization problem   $\widetilde{\Sc}_\wbf$ is upper bounded. Thus, the alternating procedure is guaranteed to converge.

\subsection{Special Case $K_i=1$}\label{sec:DL-BF}
When $K_i=1$, $\forall i$, there is only one user in each group. Problem  $\Pc_o$  is reduced to  the RIS-assisted unicast  scenario, where the BS transmits dedicated  data to each of  $G$ users 


For this unicast scenario, the RIS passive beamforming problem $\Sc_\wbf$ is still a multicast beamforming MMF problem as  in Section~\ref{subsec:RIS_problem}, which  can be efficiently solved by the proposed PSA-based algorithm in Section~\ref{Sw:PSA}.
The beamforming problem   $\Pc_\ebf$ is reduced to the conventional downlink multi-user  beamforming problem w.r.t. $\wbf$, for which a closed-form optimal solution can be obtained \cite{Bjornson&etal:SPM14,Dong&Wang:TSP2020}. In particular, SINR target for user $i$ is $\gamma_i$; for the optimal beamforming solution $\wbf_i$ in  \eqref{W_Opt}, the equivalent  channel matrix $\widetilde{\Hbf}_i$ for group $i$ reduces to an equivalent channel $\tilde{\hbf}_{i}$ for user $i$, and the weight vector $\abf_i$  becomes a scalar, denoted as $a_i$. Then, we have 
\begin{align}\label{w_opt:unicast}
\wbf_i = a_i\Rbf^{-1}(\lambdabf)\tilde{\hbf}_i, \quad i \in \Gc.
\end{align}
The weights $a_i$'s for all $G$ users can be derived in closed-form in this unicast case. This is because when the transmit beamformer can be individually designed for each user, at the optimality, all SINR constraints in \eqref{SINR_constr:mod1} can be attained with equality. Denote $\cbf_i \triangleq \Rbf^{-1}(\lambdabf)\widetilde{\hbf}_{i}$, $i \in \Gc$. Then, we have the following  SINR equality constraints 
\begin{align}\label{a_eqn}
\frac{a_i^2}{\gamma_{i}}|\cbf_i^H\hbf_{i}|^2 = \!\!\sum_{j\in \Gc, j\neq i}\! \! a_{j}^2|\cbf_{j}^H\hbf_{i}|^2 + \sigma^2, \quad \text{for}~ i \in \Gc.
\end{align} 
Solving the above $G$ linear equations, we obtain $a_i^2$, for $i\in\Gc$. In a  matrix representation, define a $G \times G$ matrix $\Vbf$ with its $ij$-th entry as
\begin{align}
\label{equ19}
[\Vbf]_{ij}= \begin{cases}
               \frac{1}{\gamma_{i}}|\cbf_i^H\hbf_{i}|^2, & j=i \\
             -|\cbf_{j}^H\hbf_{i}|^2, &  j\neq i.
   \end{cases}
\end{align}
Then, the solution to the  $G$ linear equations in \eqref{a_eqn} can be compactly written as
\begin{align}\label{UnicastSol}
\tilde{\abf}^\text{\tiny sq}=\sigma^2\Vbf^{-1}\mathbf{1}
\end{align}
where  $\tilde{\abf}^\text{\tiny sq}=[a_1^2,. . . ,a_G^2]^T$, and $\mathbf{1}$ is an all-one vector.
Finally, without the loss of optimality, we have 
$a_i=\sqrt{[\tilde{\abf}]_i}$, $i\in \Gc$.
 
In summary, in this special unicast case, the AMBF approach for solving the joint optimization of $(\wbf,\ebf)$ for $\Pc_o$   involves alternating between computing the closed-form solution $\wbf_i$ in \eqref{w_opt:unicast} and solving $\widetilde{\Sc}_{\wbf}$ for $\ebf$ using the updates in \eqref{PSA:y} and \eqref{PSA:e}, which is a computationally efficient solution.

\section{RIS-Assisted Multicast  MMF Problem}\label{sec:MMF}
Another commonly considered multicast beamforming formulation is the weighted MMF problem.  The objective is to maximize the minimum weighted SINR among users, subject to the BS transmit power budget and the unit-modulus constraint for the RIS elements. This joint optimization problem w.r.t. $\wbf$ and $\ebf$ is formulated as 
\begin{subequations}
\begin{align}
\Sc_o: \max_{\wbf, \ebf}&\min_{i, k}
 \ \ \frac{\SINR_{ik}}{\gamma_{ik}} \label{Obj-MMF}\\
 \st & \ \ \sum_{i=1}^G \|\wbf_i\|^2\le P, \label{PowerConst-MMF}\\
& \ \ |e_m|^2=1 , \ m \in \Mc. \label{UnitModConst-MMF}
\end{align}
\end{subequations}
where $P$ is the maximum transmit power limit, and the weight $\gamma_{ik}$ is a pre-specified SINR target, which can be used to set certain fairness  among users. 

Similar to the QoS problem $\Pc_o$, the MMF problem $\Sc_o$ is  non-convex  NP-hard.
We are interested in the relationship between  $\Pc_o$ and  $\Sc_o$. Note that the RIS-assisted multicast beamforming design includes optimization of the RIS passive beamformer in addition to the BS transmit beamformers. We  show below that these two problems are inverse problems.

\begin{proposition}\label{prop1}
For RIS-assisted multicast beamforming, the QoS problem $\Pc_o$ and the MMF problem $\Sc_o$ are inverse problems. Specifically,  let $\gammabf$ be the  vector containing all $\gamma_{ik}$'s. Parameterize $\Sc_o$ as $\Sc_o(\gammabf,P)$ for given $\gammabf$ and $P$ with the maximum objective value denoted as $t^o=\Sc_o(\gammabf,P)$, and
parameterize $\Pc_o$ as $\Pc_o(\gammabf)$ with $\gammabf$ being the SINR target vector, with the minimum power objective value as $P = \Pc_o(\gammabf)$. Then, $\Pc_o(\gammabf)$ and $\Sc_o(\gammabf,P)$ have the following relations:
\begin{align}\label{inv_prob2}
t^o &= \Sc_o(\gammabf, \Pc_o(t^o\gammabf)), \quad P = \Pc_o(\Sc_o(\gammabf,P)\gammabf).
\end{align} 
\end{proposition}
\emph{Proof:} See Appendix~\ref{appA}.
\vspace{0.5em}


Based on Proposition~\ref{prop1}, we can solve the MMF problem $\Sc_o$ using Algorithm 1. In particular, by the relation in \eqref{inv_prob2}, $\Sc_o$ can be solved by solving its inverse QoS problem $\Pc_o$ iteratively along with a bi-section search for $t^o$.  However, since the size of the RIS reflection elements $M$ is large, this indirect approach by iteratively solving   $\Pc_o$ can be computationally inefficient. 

In the following, we propose a more efficient algorithm to solve  $\Sc_o$ directly.
In Section~\ref{subsec:S_w}, PSA is proposed to obtain  $\ebf$ for the RIS multicast beamforming MMF problem.
Below, we propose an alternating PSA (APSA) approach to  directly compute a solution for $\Sc_o$. 
\subsection{ Proposed Fast  Algorithm} \label{subsec:MMF_alg}
Since problems $\Pc_o$ and $\Sc_o$ are the inverse problems, it implies that the optimal structure of the BS multicast beamformer $\wbf$  still has the form  given in \eqref{W_Opt}. Indeed, using the equivalent channel $\tilde{\hbf}_{ik}$ between the BS and the user, the BS mutlicast beamforming is the same as the conventional downlink multicast MMF problem, for which the optimal beamforming structure is   \cite{Dong&Wang:TSP2020}
\begin{align}
\wbf_i  =  \widetilde{\Rbf}^{-1}(\lambdabf)\widetilde{\Hbf}_i\abf_i
\end{align}
 where $\widetilde{\Rbf}=  
 {\bf{I}}+\frac{P}{\sigma^2}\sum_{i=1}^{G}\sum_{k=1}^{K}\frac{\lambda_{ik}\gamma_{ik}}{\lambdabf^T\gammabf}\tilde{\hbf}_{ik}\tilde{\hbf}_{ik}$. 

However, computing $\lambdabf$ in $\widetilde{\Rbf}(\lambdabf)$ is not straightforward for the MMF problem. Fortunately, the asymptotic expression of $\widetilde{\Rbf}(\lambdabf)$ for large $N$ has been obtained in closed-form   \cite{Dong&Wang:TSP2020}. We can use it as an approximate expression of  $\widetilde{\Rbf}(\lambdabf)$ to further simplify the computation.    
Specifically, we rewrite $\tilde{\hbf}_{iK} = \sqrt{\tilde{\beta}_{ik}}\gbf_{ik}$ where $\tilde{\beta}_{ik}$ represents\ the large-scale channel variance and  $\tilde{\gbf}_{ik}\sim \Cc\Nc(\mathbf{0}, \Ibf)$. Then, the asymptotic expression of  $\widetilde{\Rbf}(\lambdabf)$ has the following simple closed form    
\begin{align}\label{R_Approximate}
\widetilde{\Rbf}(\lambdabf) \approx \Ibf + \frac{P\bar{\beta}_{\hbf}}{\sigma^2 K_{\tot}} \sum_{i=1}^G\sum_{k=1}^{K_i}\gbf_{ik}\gbf_{ik}^H \triangleq \widetilde{\Rbf}^\text{\tiny asym}
\end{align} 
where $ \bar{\beta}_{\hbf} \triangleq 1/\Big( \frac{1}{K_{\tot}}\sum_{i=1}^G\sum_{k=1}^{K_i}\frac{1}{\tilde{\beta}_{ik}}\Big)$ is the harmonic mean of the large-scale channel variance of all users. Using $\widetilde{\Rbf}(\lambdabf)$ in \eqref{R_Approximate}, we can replace $\wbf_i$ in  \eqref{W_Opt} with the following expression:
\begin{align}
\label{W_Opt_R}
\wbf_i  \approx  (\widetilde{\Rbf}^{\text{\tiny asym}})^{-1} \widetilde{\Hbf}_i\abf_i \, , \; i \in \Gc.
\end{align}

We use $\wbf_i$ in \eqref{W_Opt_R} to develop a low-complexity fast algorithm for $\Sc_o$.
We  first convert $\Sc_o$ into a joint optimization problem w.r.t. $\{\abf_i\}$ and $\ebf$, given by
\begin{subequations}\label{eq-MMF}
\begin{align}
\Sc_o^\text{eq}: \max_{\abf, \ebf}&\min_{i, k}
 \ \ \frac{\SINR_{ik}}{\gamma_{ik}} \label{Obj-eq-MMF}\\
 \st & \ \ \sum_{i=1}^G \|\widetilde{\Cbf}_i\abf_i\|^2\le P, \label{PowerConst-eq-MMF}\\
& \ \ |e_m|^2 \leq 1 , \ m \in \Mc \label{UnitModConst_Lower-eq-MMF}\\
& \ \ \ebf^T\ebf=M. \label{UnitModConst_Upper-eq-MMF}
\end{align}
\end{subequations}
where  $\widetilde{\Cbf}_i \triangleq (\widetilde{\Rbf}^{\text{\tiny asym}})^{-1}\widetilde{\Hbf}_i$,  $\abf \triangleq [\abf_1^H, \cdots, \abf_G^H]^H$, and  the constraints in \eqref{UnitModConst-MMF} are replaced with the equivalent set of constraints in \eqref{UnitModConst_Lower-eq-MMF} and \eqref{UnitModConst_Upper-eq-MMF} based on \eqref{Unit_v:eq}.
Using the similar relaxation technique  from $\Sc_{\wbf}^\text{eq}$ to $\widetilde{\Sc}_{\wbf}$ in \eqref{v_MMF:0}, we relax $S_o^\text{eq}$ into the following problem by moving the equality constraint in \eqref{UnitModConst_Upper-eq-MMF} into the objective function as a penalty term with a penalty weight $\tilde{\rho} > 0$:
\begin{subequations}\label{MMF-relaxed}
\begin{align}
\widetilde{\Sc}_o: \max_{\abf, \ebf}& \ \bigg(\!\min_{i, k}
 \ \ \frac{\SINR_{ik}}{\gamma_{ik}}\bigg) + \tilde{\rho}\frac{\ebf^H\ebf}{M} \label{Relax-Obj-MMF}\\
 \st & \ \ \sum_{i=1}^G \|\widetilde{\Cbf}_i\abf_i\|^2\le P, \label{Relax-PowerConst-MMF}\\
& \ \ |e_m|^2 \leq 1 , \ m \in \Mc. \label{Relax-UnitModConst-MMF}
\end{align}
\end{subequations}
We note that problem  $\widetilde{\Sc}_o$ is  structurally similar to $\widetilde{\Sc}_\wbf$ in \eqref{v_MMF:0}, except that   $\widetilde{\Sc}_o$ is  a joint optimization problem for $(\abf,\ebf)$ while  $\widetilde{\Sc}_\wbf$ is for $\ebf$ only.   Thus,  we propose an APSA approach to   alternatingly apply PSA to   compute a  solution $(\abf,\ebf)$ for $\widetilde{\Sc}_o$.

\subsubsection{ Reformulation} 
We first convert $\widetilde{\Sc}_o$ into a min-max problem.
Substituting the SINR expression in \eqref{eqn:SINR} into the objective function in \eqref{Relax-Obj-MMF}, for each $k\in \Kc_i$, $i \in \Gc$, we define  
\begin{align} \label{varphi_MMF}
\hspace*{-1em}\varphi_{ik}(\abf,\ebf) \triangleq &-\!\frac{1}{\gamma_{ik}} \frac{|(\hbf_{ik}^\d \!\!+\! \Gbf_{ik}\ebf)^H\widetilde{\Cbf}_i\abf_i|^2}{\sum_{j\neq i}|(\hbf_{ik}^\d \!\!+\! \Gbf_{ik}\ebf)^H\widetilde{\Cbf}_j\abf_j|^2\!+\!\sigma^2} - \tilde{\rho}\frac{\ebf^H\ebf}{M}, 
 \end{align}
 Following the similar arguments from \eqref{phi_RIS_MMF}--\eqref{f_MMF}, let  $\varphibf(\abf,\ebf)$ be a $K_\tot\times 1$ vector containing all  $\varphi_{ik}(\abf,\ebf)$'s, and let $\ybf$   denote a $K_\tot\times 1$ vector with $\ybf\succcurlyeq 0$ and ${\bf 1}^T\ybf=1$. Then, $\widetilde{\Sc}_o$ can be  transform into the following equivalent problem
 \begin{align}
\label{g_MMF}
\widetilde{\Sc}_o':\min_{\abf \in \Ac, \ebf\in \Ec}\max_{\ybf \in\Yc}& \ \tilde{f}(\abf,\ebf,\ybf)
\end{align}
where $\tilde{f}(\abf,\ebf,\ybf) \triangleq \varphibf^T(\abf,\ebf) \ybf$, $\Ac \triangleq \{\abf:  \sum_{i=1}^G \|\widetilde{\Cbf}_i\abf_i\|^2\le P\}$,    $\Yc = \{\ybf: \ybf \succcurlyeq 0, \mathbf{1}^T\ybf = 1\}$, and $\Ec = \{\ebf: |e_m|^2 \le 1, \ m  \in \Mc\}$. All $\Ac$, $\Ec$, and $\Yc$ are compact convex sets.

\subsubsection{ Alternating Projected Subgradient Algorithm}
 Similar to
the updating steps 
in  \eqref{PSA:y} and \eqref{PSA:e} to solve  $\widetilde{\Sc}_\wbf'$ in \eqref{f_MMF}, we  can apply PSA to update $\abf$ and $\ebf$ alternatingly. In particular, the updating procedure at iteration $n$ is given by
\begin{align}
  & \ybf^{(n)} \in \displaystyle \mathop{\arg\max}_{\ybf \in \Yc} \tilde{f}(\abf^{(n)}, \ebf^{(n)}, \ybf); \label{MMFPSA:y}\\
 & \ebf^{(n+1)} = \Pi_{\Ec}\big(\ebf^{(n)} - \tilde{\alpha}\nabla_{\ebf}\tilde{f}(\abf^{(n)}, \ebf^{(n)}, \ybf^{(n)})\big) \label{MMFPSA:e}\\
&\tilde{\ybf}^{(n)} \in \displaystyle \mathop{\arg\max}_{\ybf \in \Yc} \tilde{f}(\abf^{(n)}, \ebf^{(n+1)}, \ybf);\\
 & \abf^{(n+1)} = \Pi_{\Ac}\big(\abf^{(n)} - \tilde{\alpha}\nabla_{\abf}\tilde{f}(\abf^{(n)}, \ebf^{(n+1)}, \tilde{\ybf}^{(n)})\big) \label{MMFPSA:a}
\end{align}
where $\tilde{\alpha} > 0$ is the step size, $\Pi_{\Ec}(\ebf)$ is the projection of $\ebf$ onto  $\Ec$ as in \eqref{eqn:Pi_X}, and $\Pi_{\Ac}(\xbf)$ is the projection of $\abf$ onto the feasible set $\Ac$, given by 
\begin{align}\label{eqn:U_X2}
\Pi_{\Ac}(\xbf)  =   \begin{cases} 
 \abf  & \mbox{if } \sum_{i=1}^G \|\widetilde{\Cbf}_i\abf_i\|^2\le P \\
        \sqrt{\frac{P}{P_{\text{tot}}}}\abf & \text{o.w.} 
        \end{cases}. 
\end{align}
 
Since $\Yc$ is a probability simplex,  an optimal solution to the maximization problem in \eqref{MMFPSA:y} is $\ybf = [0, \cdots, 1, \cdots, 0]^T$, with 1 at the position  that corresponds to $\max_{i,k}\varphi_{ik}(\ebf)$, \ie 
\begin{align} \label{MMFPSA:y_2}
\tilde{f}(\abf,\ebf,\ybf^{(n)})=\varphi_{\hat{i} \hat{k}} (\abf,\ebf)=\max_{i,k}\varphi_{ik}(\abf,\ebf),
\end{align}
for some pair $(\hat{i},\hat{k})$. 
Thus, the gradient $\nabla_{\xbf}\tilde{f}(\abf,\ebf, \ybf^{(n)})$, for $\xbf=\abf$ or $\ebf$, is given by
\begin{align}
\nabla_{\xbf}\tilde{f}(\abf,\ebf, \ybf^{(n)}) &=\! \nabla_{\xbf}\varphi_{\hat{i} \hat{k}} (\abf,\ebf) , \quad \text{for~} \xbf=\abf \text{~or~} \ebf.   \label{Gradient_g(x)}
\end{align}

For  $\nabla_{\ebf}\varphi_{\hat{i} \hat{k}} (\xbf)$, note that the expression for  $\varphi_{ik}(\xbf)$ is the same as $\phi_{ik}(\ebf)$. Thus, the gradient $\nabla_{\ebf}\varphi_{\hat{i} \hat{k}} (\xbf)$ can be computed using \eqref{PSA:f_grad1}, where  $\wbf_i$ is computed using \eqref{W_Opt_R}, and $\rho$ is replaced by  $\tilde{\rho}$.

To compute $\nabla_{\abf}\varphi_{\hat{i} \hat{k}} (\abf,\ebf)$,  we rewrite $\varphi_{ik}(\abf,\ebf)$ in \eqref{varphi_MMF} as
\begin{align*}
\varphi_{ik}(\abf,\ebf) = &-\frac{1}{\gamma_{ik}} \frac{\abf_i^H\widetilde{\Abf}_{iik}\abf_i}{\sum_{j\neq i}\abf_j^H\widetilde{\Abf}_{jik}\abf_j+\sigma^2} - \tilde{\rho}\frac{\ebf^H\ebf}{M}, 
\end{align*}
where $\widetilde{\Abf}_{jik} \triangleq \widetilde{\Cbf}_j^H\tilde{\hbf}_{ik}\tilde{\hbf}_{ik}^H\widetilde{\Cbf}_j$. Then, we have
\begin{align}\label{PSA:g_grad1}
 \nabla_{\abf}\varphi_{\hat{i} \hat{k}} (\abf,\ebf) &  =  [\nabla_{\abf_1}^H\varphi_{\hat{i} \hat{k}} (\abf,\ebf), \cdots, \nabla_{\abf_G}^H\varphi_{\hat{i} \hat{k}}(\abf,\ebf)]^H
\end{align}
where
\begin{align}\label{PSA:g_grad2}
\hspace*{-.8em} \nabla_{\abf_i}\varphi_{\hat{i} \hat{k}} (\abf,\ebf) \!= \!\begin{cases}
   \displaystyle -\frac{1}{\gamma_{\hat{i}\hat{k}}} \frac{\widetilde{\Abf}_{\hat{i}\hat{i} \hat{k}}\abf_{\hat{i}}}{\sum_{j\neq \hat{i}}\abf_j^H\widetilde{\Abf}_{j\hat{i}\hat{k}}\abf_j\!+\!\sigma^2} & \mbox{if } i = \hat{i} \\[1.5em]
  \displaystyle  \frac{1}{\gamma_{\hat{i}\hat{k}}} \frac{(\abf_{\hat{i}}^H\widetilde{\Abf}_{\hat{i}\hat{i} \hat{k}}\abf_{\hat{i}})\widetilde{\Abf}_{i\hat{i}\hat{k}}\abf_i}{\bigl(\sum_{j\neq \hat{i}}\abf_j^H\widetilde{\Abf}_{j\hat{i}\hat{k}}\abf_j\!+\!\sigma^2\bigl)^2} & \text{o.w.}
\end{cases}
\end{align}

Thus, the updates in \eqref{MMFPSA:y}--\eqref{MMFPSA:a} can be computed using closed-form evaluations from  \eqref{eqn:U_X2}--\eqref{PSA:g_grad2}.

\subsubsection{Final Processing} Recall that $\widetilde{\Sc}_o$ is a relaxed problem of $\Sc_o$. Thus, the computed solution  $\ebf^\star$ may not satisfy \eqref{UnitModConst-MMF} for  $\Sc_o$. We use this final processing step to obtain a feasible solution to $\Sc_o$. Let   $({\abf^\star}, {\ebf^\star})$ be the solution produced by the PSA procedure  in \eqref{MMFPSA:y}--\eqref{MMFPSA:a}. If $\ebf^\star$ does not satisfy \eqref{UnitModConst-MMF}, we have the following step to obtain a feasible solution based on $(\abf^\star,\ebf^\star)$:
\begin{itemize}
\item Project $\ebf^\star$ onto its feasible set: $\ebf^{\text{final}} = \exp(j\angle \ebf^\star)$, which obtains the phase component of each element in $\ebf^\star$;
\item Given $\ebf^{\text{final}} $, we solve  $\Sc_o^{\text{eq}}$ in \eqref{eq-MMF} w.r.t. $\abf$ to obtain the final weight vector $\abf^{\text{final}}$:
\begin{align}\label{finalproc_MMF}
\Sc^\text{eq}(\ebf^{\text{final}}): & \max_{\abf}\min_{i, k}
  \frac{\SINR_{ik}}{\gamma_{ik}} \nn \\
& \st  \sum_{i=1}^G \|\widetilde{\Cbf}_i\abf_i\|^2\le P 
\end{align}
where $\SINR_{ik}$ in \eqref{eqn:SINR} is computed based on  $\ebf^{\text{final}}$. This is the conventional MMF problem, which can by solved by the PSA-based fast algorithm \cite{Zhang&etal:COML22}.  Note that  $\abf^\star$ satisfies the power constraint in \eqref{finalproc_MMF} and thus serves as a good initial point to this algorithm to achieve fast convergence and good performance. The final beamforming solution $\wbf^{\text{final}}$ is computed using \eqref{W_Opt_R} based on  $\abf^{\text{final}}$. 
\end{itemize}

We summarize our proposed fast algorithm  for the   MMF problem $\Sc_o$ in Algorithm \ref{alg2}. Again, similar to that discussed in Section~\ref{subsec:conv}, we track the best solution for $(\abf,\ebf)$ with  the minimum objective value of $\widetilde{\Sc}_o'$ obtained so far in $n$ iterations: $\tilde{g}_{\min}^{(n)}=\min\{\tilde{g}^{(1)},...,\tilde{g}^{(n)}\}$, where $\tilde{g}^{(n)}=\max_{\ybf \in\Yc} f(\abf^{(n)},\ebf^{(n)},\ybf)$. Since, $\tilde{g}_{\min}^{(j)}$ is monotonically decreasing and is bounded below,  we terminate the algorithm when $\tilde{g}_{\min}^{(j)}$ converges. 
\begin{algorithm}[t]
\caption{The  APSA  Algorithm for RIS-Assisted MMF Problem $\Sc_o$.}\label{alg2}
\begin{algorithmic}[1]
\State  \textbf{Initialization:} Set  feasible point $(\abf^{(0)},\ebf^{(0)})$; Set  $\tilde{\rho}$ and  $\tilde{\alpha}$; Set $n=0$.
\Repeat
\State  Compute $\varphi_{ik}(\abf^{(n)}, \ebf^{(n)})$, $k\in\Kc_i$, $i\in\Gc$.
Obtain pair $(\hat{i},\hat{k})$ as in \eqref{MMFPSA:y_2}.
\State Update $\ebf^{(n+1)}$ using \eqref{PSA:f_grad1} and  \eqref{eqn:Pi_X}.
\State  Compute $\varphi_{ik}(\abf^{(n)}, \ebf^{(n+1)})$, $k\in\Kc_i$, $i\in\Gc$.
Obtain pair $(\hat{i},\hat{k})$ as in \eqref{MMFPSA:y_2}.
\State Update  $\abf^{(n+1)}$ using  \eqref{PSA:g_grad2} and \eqref{eqn:U_X2}.
\State  Set $n \gets n+1$.
\Until convergence
\State Set  $\ebf^{\text{final}}=\exp(j\angle\ebf^{(n)})$. 
\State Solve $\Sc^\text{eq}(\ebf^{\text{final}})$ to obtain $\abf^{\text{final}}$.
\State Compute $\wbf^{\text{final}}$ using \eqref{W_Opt_R}.
\State \textbf{return} ($\wbf^{\text{final}}$, $\ebf^{\text{final}}$).
\end{algorithmic}
\end{algorithm}

\subsection{ Algorithm Discussion}
 
\subsubsection{Initialization} The proposed updating procedure requires an initial point  $({\abf^{(0)}}, {\ebf^{(0)}})$. This initial point can be generated randomly, as the projections $\Pi_{\Ec}$ and $\Pi_{Ac}$  will ensure that the subsequent points $\{({\abf^{(n)}}, {\ebf^{(n)}})\}$ are feasible. However,  a good initial point help accelerate the convergence. Thus, instead of a random initial point, we consider the following approach. 

We first generate  $\ebf^{(0)}$  with  a random phase for each element.  Given $\ebf^{(0)}$,  $\Sc_o^\text{eq}$ in \eqref{eq-MMF} is reduced to the  MMF problem w.r.t. $\abf$, \ie $\Sc^\text{eq}(\ebf^{(0)})$ as in \eqref{finalproc_MMF} by replacing $\ebf^{\text{final}}$ with $\ebf^{0}$. Note that any feasible solution $\abf$ to $\Sc^\text{eq}(\ebf^{(0)})$  is feasible to $\widetilde{\Sc}_o$. Thus, we only need to find a feasible solution. To do so, we solve the inverse problem of $\Sc^\text{eq}(\ebf^{(0)})$, \ie the QoS problem similar to $\Pc_\ebf$ in \eqref{Pe} for a given minimum SINR\ target $t$,  along with one round bi-section search over $t$, to find the transmit power objective of the  QoS problem\footnote{To solve the QoS problem, we can use the classical approach of SDR with the Gaussian randomization procedure.}  that is less than $P$ and obtain the initial point $\abf^{(0)}$.

\subsubsection{Computational Complexity} For the updates in \eqref{MMFPSA:y}--\eqref{MMFPSA:a} at each iteration, the update $\ybf^{(j)}$ is obtained in \eqref{MMFPSA:y_2} by finding the maximum $\varphi_{ik}(\ebf)$ among $K_\tot$ users, which requires $GK_\tot (N+1)+K_\tot(MN+2N+4)$ flops. The updates in \eqref{MMFPSA:e} and \eqref{MMFPSA:a} require to calculate $\nabla_{\abf}\varphi_{\hat{i} \hat{k}} (\xbf)$ and $\nabla_{\ebf}\varphi_{\hat{i} \hat{k}}(\xbf)$. Note that $\nabla_{\ebf}\varphi_{\hat{i} \hat{k}} (\xbf)$ is computed using  \eqref{PSA:f_grad1}, which  requires $2GM + 2M$ flops as discussed  in Section~\ref{subsec:complexity_QoS}.ii). To compute $\nabla_{\abf_i}\varphi_{\hat{i} \hat{k}}(\xbf)$ in \eqref{PSA:g_grad2}, we need   
$\abf_j^H\widetilde{\Cbf}_j^H\tilde{\hbf}_{\hat{i}\hat{k}}$ for $j\in \Gc$, which is part of the SINR\ expression that has already been computed in $\varphi_{ik}(\ebf)$. We also need to calculate $\widetilde{\Cbf}_{i}^H\tilde{\hbf}_{\hat{i}\hat{k}}$ for  $\widetilde{\Abf}_{i\hat{i}\hat{k}}\abf_{i}=\widetilde{\Cbf}_{i}^H\tilde{\hbf}_{\hat{i}\hat{k}}\tilde{\hbf}_{\hat{i}\hat{k}}^H\widetilde{\Cbf}_{i}\abf_{i}$, for $i\in \Gc$.
Note that $\widetilde{\Cbf}_i$'s only need to  be computed once before the update procedure, and $\tilde{\hbf}_{\hat{i}\hat{k}}$ is already computed as part of the SINR expression. Thus, compute   $\widetilde{\Abf}_{i\hat{i}\hat{k}}\abf_{i}$'s for $i\in \Gc$ require $K_\tot (N+1)$.
Lastly, we need $K_{\tot}+G$ flops to calculate the final value of $\nabla_{\abf_i}\varphi_{\hat{i}\hat{k}}(\xbf)$ for all $\abf_i$'s. In summary, the complexity in obtaining the updates in   \eqref{MMFPSA:e} and \eqref{MMFPSA:a} require $2(G+1)M+K_\tot(N+2)+G$ flops. Thus, the overall  leading complexity of the updating procedure in each PSA iteration is dominated by computing  $\ybf^{(j)}$ in \eqref{MMFPSA:y_2}, which is  $ GK_\tot N+K_\tot MN$ flops. It grows linearly with $M$ and $N$.

\subsubsection{Special Case $K_i=1$}
When $K_i=1$, $\forall i$, problem $\Sc_o$ reduces to the RIS-assisted  multiuser unicast scenario for $G$ users with the max-min fair objective.  In this case, the computational complexity per iteration in Algorithm~\ref{alg2} reduces to  $GMN+G^2N$ flops, which grows linearly with  $M$ and $N$.

\subsubsection{Improved Initialization Method for QoS Problem $\Pc_o$} We can adopt the APSA algorithm as an improved  initialization method to AMBF (Algorithm~\ref{alg1}) for the QoS problem. Specifically, following the random phase initialization  in Section~\ref{subsubsec:init_QoS}, we have the initial point $(\ebf^{0},\wbf^{0})$. We feed this point to APSA with power budget set to $P=\sum_{i=1}^G\|\wbf_i^0\|^2$.  The output $\ebf^\text{final}$ from APSA is then used as the updated initial point for $\ebf$  in Algorithm~\ref{alg1}: $\ebf^0\leftarrow \ebf^\text{final}$. Also, as the purpose of APSA here is for providing an initial point, we can set a pre-set number of iterations for Algorithm~\ref{alg2} for the output before waiting for the algorithm converges,

\section{Simulation Results} \label{sec:sim}
We consider a downlink RIS-assisted transmission scenario with a BS and an RIS. The locations of the  BS and the RIS using the $(x,y,z)$-coordinates in meters are  $(x_{\text{BS}},y_{\text{BS}},z_{\text{BS}})=(0, 0, 20)$ and  $(x_{\text{RIS}},y_{\text{RIS}},z_{\text{RIS}})=(70, 70, 20)$, respectively. The RIS is a rectangular surface with the array elements placed on the $(y, z)$-plane. The users are located randomly within a rectangle area $[45,65] \times [75,125]$ on the $(x,y)$-plane, each at location $(x_{ik},y_{ik},0)$. We set the target SINR for each user the same: $\gamma_{ik}=\gamma$, $\forall i,k$. Unless specified otherwise, our default system setup is given as $N=32$ antennas at the BS, and $M = 10 \times 10$ reflective elements at the RIS, $G=2$ groups, $K_{i}=K=3$ users, $\forall i$, and SINR target $\gamma = 10~$dB. 

We consider the system operates at carrier frequency $f_c=3.5$ GHz. We assume a line-of-sight (LOS) channel between the BS and the RIS and model $\Hbf_{\r}$ as a Rician fading channel matrix given by $\Hbf_\r = \beta^\text{BR}\big(\sqrt{\frac{K_\B}{1+K_\B}}\Hbf_\r^\text{LOS}+\sqrt{\frac{1}{1+K_\B}}\Hbf_\r^\text{NLOS}\big)$, where Rician factor   $K_\B=10$,    $\beta^\text{BR}$ is the path gain modeled as $\beta^\text{BR} \text{[dB]}=-30 - 22 \log 10(d_{\text{B-R}})$  with $d_{\text{B-R}}$ being the BS-RIS distance in meters, $\Hbf_\r^\text{NLOS}$ is the non-line-of-sight (NLOS) component modeled as  $[\Hbf_\r^\text{NLOS}]_{ij}\sim \Cc\Nc(0,\Ibf)$, and $\Hbf_\r^\text{LOS}$ is the LOS component. The LOS component $\Hbf_\r^\text{LOS}$ is a function of the BS and RIS locations and is modeled as
\begin{align}\label{Hr{LOS}}
\Hbf_\r^\text{LOS} = \bbf_{\text{RIS}}(\phi_2, \theta_2) \bbf_{\text{BS}}(\phi_1, \theta_1)^H
\end{align}
where $\phi_1$ and $\theta_1$ are the azimuth and elevation angles
of departure (AoD) from the BS to the IRS, and $\bbf_{\text{BS}}(\phi_1, \theta_1)$ is the BS antenna array 3D steering vector. Assume the BS antenna elements are along the $x$-coordinate,  the $n$th entry  of  $\bbf_{\text{BS}}(\phi_1, \theta_1)$ is given by 
\begin{align*}
[\bbf_{\text{BS}}(\phi_1, \theta_1)]_n = e^{-j\frac{2\pi(n-1)\Delta_{\text{BS}}}{\lambda_c}\cos(\phi_1)\sin(\theta_1)}, \ n =1,\ldots,N
\end{align*}
where $\Delta_{\text{BS}}$ is the distance between two adjacent BS antennas, $\lambda_c$ is the carrier wavelength, and $\cos(\phi_1)\sin(\theta_1) = \frac{x_{\text{RIS}} - x_{\text{BS}}}{d_{\text{B-R}}}$. We assume $\Delta_{\text{BS}}=\frac{\lambda_c}{2}$. Similarly,  $\bbf_{\text{RIS}}(\phi_2, \theta_2)$ is the $M\times 1$  RIS element streaming vector, with  $\phi_2$ and $\theta_2$ being the azimuth and elevation angles of arrival  from the BS to the RIS, and  the $m$th element, for $m\in \Mc$, is given by 
\begin{align}\label{b{RIS_BS}}
[\bbf_{\text{RIS}}(\phi_2, \theta_2)]_m = 
 e^{-j\frac{2\pi \Delta_{\text{RIS}}}{\lambda_c} [y_m\sin(\phi_2)\sin(\theta_2)+ z_{m}\cos(\theta_2)]}
\end{align}
where $\Delta_{\text{RIS}}=\frac{\lambda_c}{2}$ is the distance between two adjacent RIS elements, $y_m = \text{mod}(m-1, M_y)$ with $M_y$ being the number of RIS elements along the $y$-axis,  $z_m =- \lfloor(m-1)/ M_y \rfloor$, $\sin(\phi_2)\sin(\theta_2) = \frac{y_{\text{BS}} - y_{\text{RIS}}}{d_{\text{B-R}}}$,  and $\cos(\theta_2) = \frac{z_{\text{BS}} - z_{\text{RIS}}}{d_{\text{B-R}}}$.  

We assume a LOS channel between the RIS and each user.
The channel   $\hbf^{\r}_{ik}$ between the RIS and user $k$ in group $i$ is  generated i.i.d as $\hbf^{\text{r}}_{ik} = \beta^\r_{ik} \big(\sqrt{\frac{K_\r}{1+K_\r}}\hbf_{ik}^\text{r,LOS}+\sqrt{\frac{1}{1+K_\r}}\hbf_{ik}^\text{r,NLOS}\big)$, where Rician factor   $K_\r=10$,    $\beta^\r_{ik}$ is the  path gain modeled as $\beta^\r_{ik} [\text{dB}]= -30 - 22 \log 10(d^{\text{R-U}}_{ik})$, with $d^\text{R-U}_{ik}$ being the  distance between the RIS and the user. The NLOS component is $\hbf_{ik}^\text{r,NLOS} \sim \Cc\Nc({\bf 0}, \Ibf)$. The LOS component $\hbf_{ik}^\text{r,LOS}=\bbf_{\text{RIS}}(\phi_{ik}^\r, \theta^\r)$ is similar as in \eqref{b{RIS_BS}},  where  $\phi_{ik}^\r$ and $\theta^\r$ being the azimuth and elevation AoD from the IRS to each user, with 
$\sin(\phi_{ik}^\r)\sin(\theta^\r) = \frac{ y_{ik} - y_{\text{RIS}} }{d^\text{R-U}_{ik}}$,  and $\cos(\theta^\r) = \frac{z_{ik}-z_{\text{RIS}}}{d^\text{R-U}_{ik}}$. 

The channel between BS and user $k$ in group $i$   is modeled as a NLOS channel with Rayleigh fading, which is generated i.i.d as $\hbf^{\text{d}}_{ik} \sim \Cc\Nc({\bf 0}, \beta^\d_{ik}\Ibf)$, where $\beta^\d_{ik}$ is the  path gain modeled as $\beta^\d_{ik} [\text{dB}]= -32.4 - 36.7 \log 10(d^{\text{B-U}}_{ik})$, with $d^\text{B-U}_{ik}$ being the respective BS-user distances. The receiver noise power is set to $\sigma^2 = -100~\text{dBm}$. 

In Algorithm \ref{alg1} for the QoS problem, we set the penalty parameter in subproblem $\widetilde{\Sc}_\wbf$ to $\rho = 1$, and the step size for the PSA updates in \eqref{PSA:e} as $\alpha =  2.5$. We set the convergence threshold of PSA for RIS MMF subproblem $\tilde{\Sc}_\wbf$ to be $g_{\min}^{(j)}-g_{\min}^{(j-1)} \le 1e-5$, and the overall AMBF objective value convergence threshold to be $1e-3$. In  Algorithm \ref{alg2} for the MMF problem, we set the penalty parameter in $\widetilde{\Sc}_o$ to $\tilde{\rho} = 1$, and the step size $\tilde{\alpha}$ for the updates in \eqref{MMFPSA:e} and \eqref{MMFPSA:a} as $\tilde{\alpha} =  1.5$.\footnote{We have conducted experiments using different values of $(\rho, \alpha)$ and $(\tilde{\rho}, \tilde{\alpha})$. The selected parameter values in general provide the best  balance between the performance and the convergence rate.} The APSA convergence threshold is set to $\tilde{g}_{\min}^{(n)}-\tilde{g}_{\min}^{(n-1)}  \le 1e-5$.

\subsection{Performance  for the QoS Problem}
\subsubsection{Convergence Behavior}
\begin{figure}[t]
        \begin{subfigure}[b]{0.25\textwidth}
        \centering
        \includegraphics[width=\textwidth]{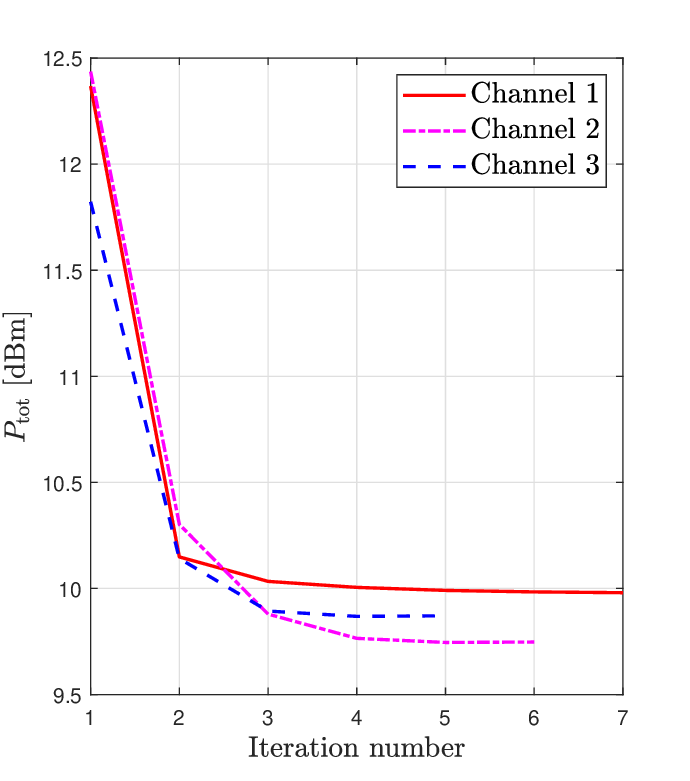}
        \caption{The overall AMBF for $\Pc_o$.}
        \label{fig:PowervsMCR}
        \end{subfigure}
        \hfill
       \hspace*{-2em} \begin{subfigure}[b]{0.25\textwidth}
        \centering
        \includegraphics[width=\textwidth]{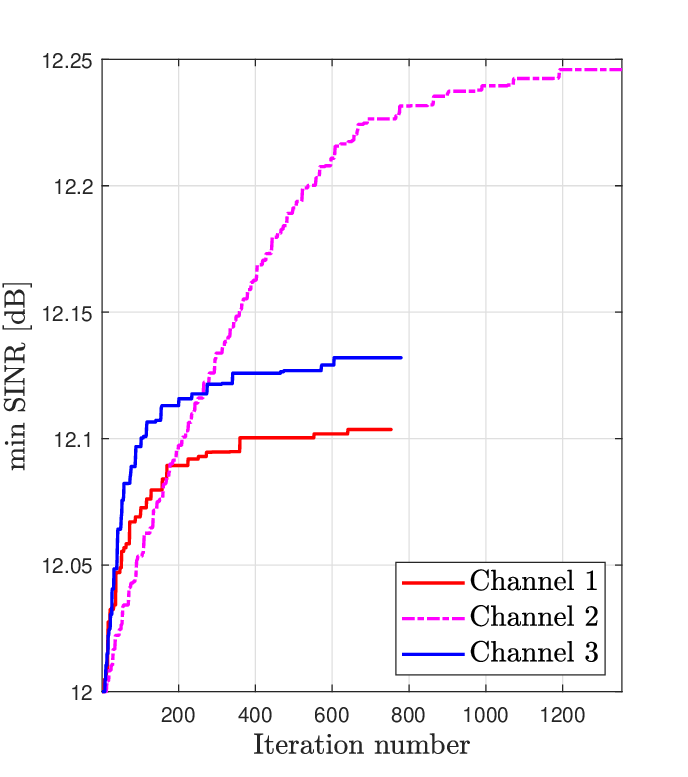}
        \caption{The subproblem $\widetilde{\Sc}_{\wbf}$.}
        \label{fig:MinSINRvsIter-QoS}
        \end{subfigure}
        \caption{\footnotesize Convergence behavior of  Algorithm~\ref{alg1} ($N=32, M=100$, $(G,K)=(2,3)$,  $\gamma = 10$ dB).}
\vspace*{-1em}
\end{figure}

We first study the convergence behavior of the proposed AMBF (Algorithm \ref{alg1}) for  $\Pc_o$.  Denote the total transmit power at the BS as  $P_{\tot} = \sum_{i=1}^G \|\wbf_i\|^2$.  Fig.~\ref{fig:PowervsMCR} shows the trajectory of  $P_\tot$ over AO iterations by Algorithm \ref{alg1} for three random channel realizations. We observe that Algorithm \ref{alg1}  converges fast, especially given that  fast algorithms are used for  each subproblem. Fig.~\ref{fig:MinSINRvsIter-QoS} shows the convergence behavior of solving subproblem $\widetilde{\Sc}_{\wbf}$ for RIS phase shift $\ebf$ using PSA in the first AO  iteration of  AMBF. The convergence behavior for three random channel realizations are shown. We see PSA converges after $500\sim1400$ iterations. Note that the convergence speed  becomes faster in the subsequent AO\ iterations as the algorithm converges. Furthermore, 
the closed-form updates for PSA in each iteration is computationally inexpensive as shown in the complexity analysis. Thus, the overall computation time is fast despite that the number of iterations can be large especially in the initial AO\ iteration. 
\subsubsection{Performance Comparison}
For performance comparison, we consider the following methods:
\begin{enumerate}
\item[i)] No RIS: A conventional downlink  multicast scenario without the RIS. 
\item[ii)]  Random RIS: Apply random phase-shift for  the RIS elements in $\ebf$. With given $\ebf$, solve the BS multicast beamforming problem  $\Pc_\ebf$  for $\wbf$. 
\item[iii)] DirectSCA \cite{Kumar:WCL22}: An SCA method to directly solve $\Pc_o$. Note that this SCA algorithm was  developed originally for the RIS-assisted joint beamforming in the multiuser unicast  scenario.  We can straightforwardly extend it to solve  $\Pc_o$ in the multicast scenario. The SINR constraints w.r.t. $\wbf$ and $\ebf$ in \eqref{SINR_constr} are convexified and iteratively solved via a convex solver such as CVX \cite{Grant&Boyd:CVX}. This method serves as a competitive benchmark.  
\item[vi)] IP Method: Apply the interior-point (IP) algorithm for the  joint optimization problem. In particular, we adopt the  MATLAB built-in non-linear optimization solver {\tt Fmincon} to solve  $\Pc_o$. 

\item[v)] DirectAO \cite{Li&etal:VTC20}: An algorithm that directly applies AO to   $\Pc_o$, where the subproblem for $\ebf$ is  a feasibility problem to satisfy the SINR\ constraints in \eqref{SINR_constr} and the unit-modulus constraints in \eqref{Unit_Mod_constr}. Each subproblem is solved by an SDR-based algorithm. 
\end{enumerate}

\begin{figure}[t]
        \centering
        \includegraphics[width=0.8\columnwidth]{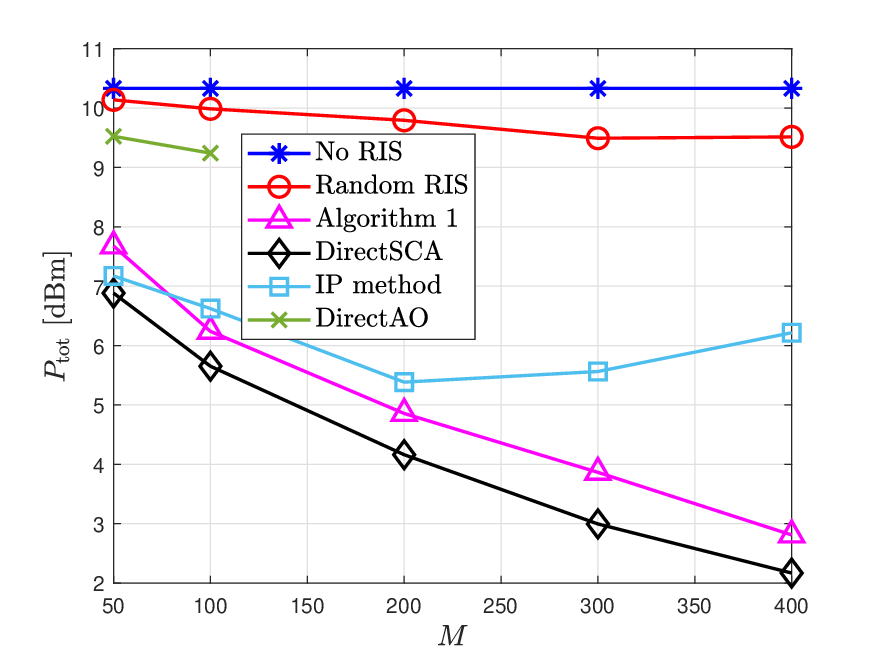}
        \caption{\footnotesize Total transmit power $P_{\tot}$  vs. $M$ ($N=32$,   $(G,K)=(2,3)$,  $\gamma = 10$ dB).}
        \label{fig:PowervsMKr0}\vspace*{-1.5em}
\end{figure}

 Fig.~\ref{fig:PowervsMKr0} shows the total BS transmit power   $P_{\tot}$ vs. the number of RIS  elements $M$ by different methods, for the  target SINR $\gamma= 10$~dB.  The performance of random RIS remains roughly flat as $M$ increases, as it does not provide an effective  RIS beamforming solution. DirectAO can only provides a small improvement over random RIS, which shows solving the feasibility problem for RIS subproblem does not give an effective solution. Also, the algorithm entails very high computation complexity, and thus its performance is only shown for $M \le 100$.    In contrast, a substantial gain can be achieved when RIS beamforming is effectively optimized. As we see, DirectSCA performs the best with the lowest power. This is expected as it is a joint optimization approach to solve $\Pc_o$ directly, and SCA performs well in general. However, the main drawback of this approach is the cost of impractically high computational complexity as we will show later.   In addition, as $M$ becomes large,  the numerical stability of this method is reduced noticeably. In particular, the frequency that the CVX solver fails to find a solution  increases to $40\%\sim 50\%$ when $M>100$.  In comparison, the performance of the proposed Algorithm~\ref{alg1} closely follows that of DirectSCA, with a gap about $1$~dB. Also, the algorithm always provides a converged  solution. IP method  for this non-convex problem often has trouble to converge properly, and  the resulting performance  deteriorates as $M$ increases, with  a large gap observed for $M>200$.  
\begin{table}[t]
\renewcommand{\arraystretch}{1.2}
\centering
\caption{Average Computation Time (sec).}
\label{tab:vsM}
\vspace*{-.5em}
\begin{tabular} {p{1.85cm}||p{0.6cm}|p{0.6cm}|p{0.6cm}|p{0.6cm}|p{.7cm}}
\hline
$M$ & 50 & 100 & 200 & 300 & 400\\
\hline\hline
Algorithm~\ref{alg1}  & 6.49   & 17.03 & 22.2& 24.8 & 37.5\\
\hline
DirectSCA \cite{Kumar:WCL22} & 646 & 935 & 2364 & 2425 & 4403\\
\hline 
IP Method & 15.9 & 19.9 & 35.1 & 39.1 & 47.7\\
\hline 
DirectAO & 269 & 1015 & ~~--  &~~--  &~~ --\\
\hline
\end{tabular}\vspace*{-.5em}
\end{table}

We now compare the average computation time of different  algorithms   in Table~\ref{tab:vsM}. The computational advantage of Algorithm~\ref{alg1} over DirectSCA, IP Method, and DirectAO  is clearly observed as  $M$ increases, demonstrating the excellent scalability of Algorithm~\ref{alg1} over $M$. DirectSCA has prohibitively high computational complexity as $M$ increases, making it impractical for real implementation. In contrast, the computational time of Algorithm~\ref{alg1} is several magnitudes lower and only grows mildly with $M$.  Thus, Algorithm~\ref{alg1} is a much more efficient algorithm offering competitive performance with significantly lower computational complexity as compared to DirectSCA, especially for the large value of $M$.

\begin{figure}[t]
        \centering
        \includegraphics[width=.8\columnwidth]{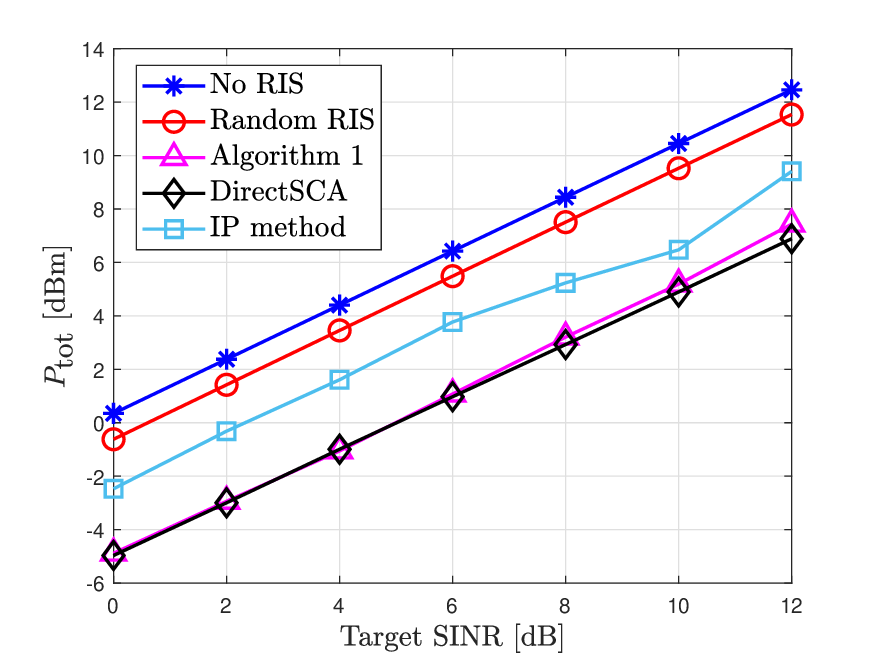}
        \caption{\footnotesize Total transmit power $P_{\tot}$  vs. target SINR $\gamma$ ($N=32, M=256$, $(G,K)=(2,3)$).}
        \label{fig:PowervsSINRKr0}
        \vspace{-1.5em}
\end{figure}

Fig.~\ref{fig:PowervsSINRKr0} shows $P_{\tot}$ vs. target SINR $\gamma$ by different methods, for $M=256~ (16\times 16)$. We see that   the proposed Algorithm~\ref{alg1} closely follows DirectSCA and provides substantial power reduction over   Random RIS and IP method  for all values of $\gamma$, with a roughly constant gap.

\subsection{Performance for the MMF Problem}
\subsubsection{Convergence Behavior}
\begin{figure}[t]
        \centering
        \includegraphics[width=0.85\columnwidth]{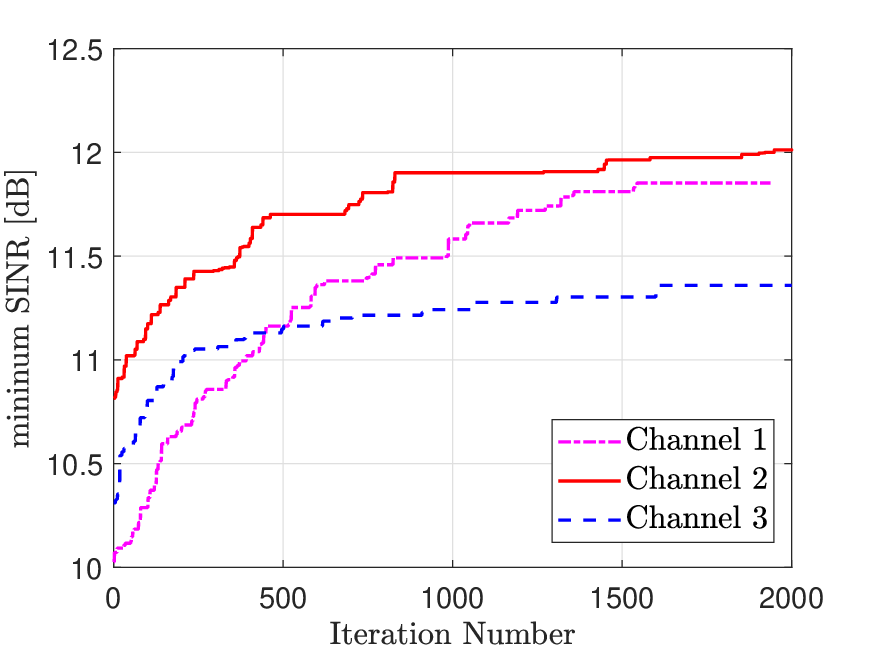}
        \centering
        \caption{\footnotesize Convergence behavior of the Algorithm \ref{alg2} ($N=32$, $M=100$, $(G,K)=(2,3)$).}
        \label{fig:MinSINRvsIter-MMF}\vspace*{-1em}
\end{figure}

Fig.~\ref{fig:MinSINRvsIter-MMF} shows the convergence behavior of Algorithm \ref{alg2} for the MMF problem $\Sc_o$   under three random channel realizations.  We see  that APSA converges in $1000\sim2000$ iterations. Since 
the closed-form updates for APSA in each iteration is computationally inexpensive, the overall computation time is fast despite the number of iterations is large. As we will see later, the overall computational time is low.

\subsubsection{Performance Comparison}

For performance comparison, besides our proposed Algorithm~\ref{alg2}, we consider the following methods:
\begin{enumerate}
\item No RIS: A conventional downlink multi-group multicast scenario without RIS. 
\item  Random RIS: Apply random phase-shift for  the RIS elements in $\ebf$. With given $\ebf$, solve the BS multicast beamforming MMF problem  for $\wbf$.
\item QoS2MMF: Solving $\Sc_o$ by iteratively solving $\Pc_o$ as discussed below Proposition~\ref{prop1} in Section~\ref{sec:MMF}.
\item IP Method: Adopt the IP algorithm as a joint optimization method to directly solve  the  MMF problem $\Sc_o$ by using the MATLAB non-linear optimization solver {\tt Fmincon}.
\end{enumerate}
 \begin{figure}[t]
        \centering
        \begin{psfrags}
        \centering
        \includegraphics[width=.8\columnwidth]{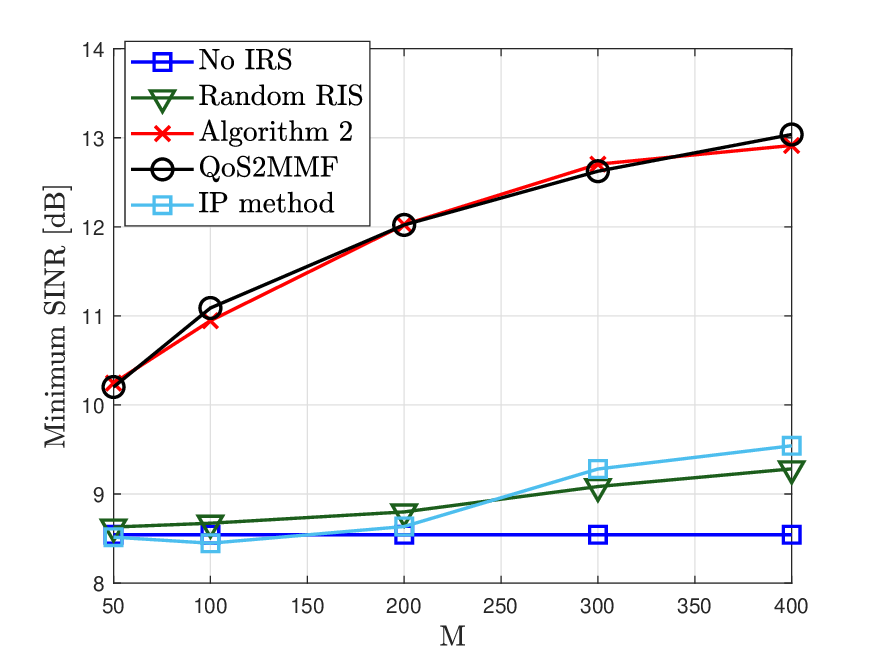}\\[-.5em]
        \end{psfrags}
        \caption{\footnotesize Average minimum SINR  vs. $M$ ($N=32$, $(G,K)=(2,3)$).}
        \label{fig:MinSINRvsM}
        \vspace{-.5em}
\end{figure}

\begin{table}[t]
\renewcommand{\arraystretch}{1.2}
\centering
\caption{Average Computation Time (sec) ($N=32$, $(G,K)=(2,3)$).}
\label{tab:MMFvsM}
\begin{tabular} {p{1.3cm}||p{0.6cm}|p{0.6cm}|p{0.6cm}|p{0.6cm}|p{.7cm}}
\hline
$M$ & 50 & 100 & 200 & 300 & 400\\
\hline\hline
Algorithm~\ref{alg1}  & 1.19    & 1.22 & 1.64 & 1.75 & 1.96\\
\hline
QoS2MMF & 27.4 & 32.5 & 37.8 & 34.1 & 58.4\\
\hline 
IP Method & 7.47 & 9.84 & 15.3 & 22.2 & 27.5\\
\hline 
\end{tabular}\vspace*{-1em}
\end{table}

Note that we select the above methods for comparison  as there are no existing methods available for the  RIS-assisted MMF problem,\footnote{ The sum-group-rate maximization problem is considered as the objective in \cite{Zhou&etal:TSP20,Farooq&etal:Globecom22}, where the group rate is defined as the minimum rate (\ie SINR) within the group. Since this objective cannot ensure fairness among groups, the solutions can result in some group having nearly zero rate, which is undesirable.} nor  a tight upper bound for meaningful comparison. In the following simulation results, we set the BS transmit power $P = 10~{\text{dBm and $\gamma_{ik}=\gamma = 10$~dB.}}$.

Fig.~\ref{fig:MinSINRvsM} shows the average minimum SINR vs.  the number of RIS elements $M$ by different methods. We see that the minimum SINR achieved by Algorithm~\ref{alg2} is substantially higher than random RIS and IP method. Also, the SINR gain under Algorithm~\ref{alg2} by increasing $M$ from $50$ to $500$ is nearly 3~dB,  significantly more than random RIS and IP method. This shows that Algorithm~\ref{alg2} can provide an effective beamforming solution for RIS elements to improve the performance. Also, the performance of  Algorithm~\ref{alg2}  and QoS2MMF are close, demonstrating that  Algorithm~\ref{alg2} is an effective first-order algorithm that provides a good performance. In contrast, IP method  cannot produce an effective solution for the MMF problem and its performance is similar to random RIS. We compare the average computation times of these methods   in Table~\ref{tab:MMFvsM}. The computational advantage of Algorithm~\ref{alg2} over  QoS2MMF and IP method can be clearly seen. 

\begin{figure}[t]
        \centering
        \includegraphics[width=.8\columnwidth]{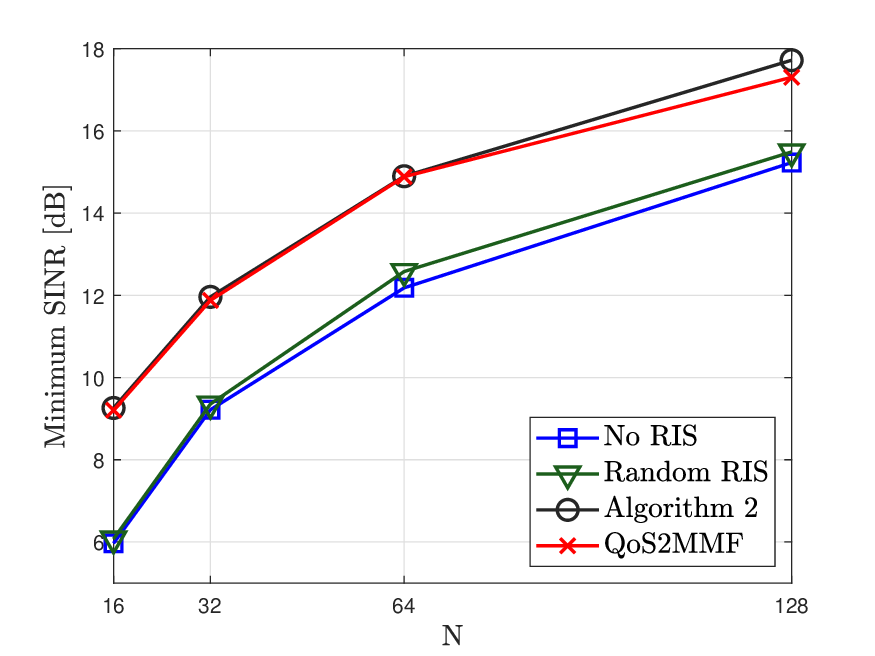}\\[-.5em]
        \centering
        \caption{\footnotesize Average minimum SINR  vs. $N$ ($M=100$, $(G,K)=(2,3)$).}
        \label{fig:MinSINRvsN}
\vspace{-1em}
\end{figure}

\begin{figure}[t]
        \centering
        \includegraphics[width=.8\columnwidth]{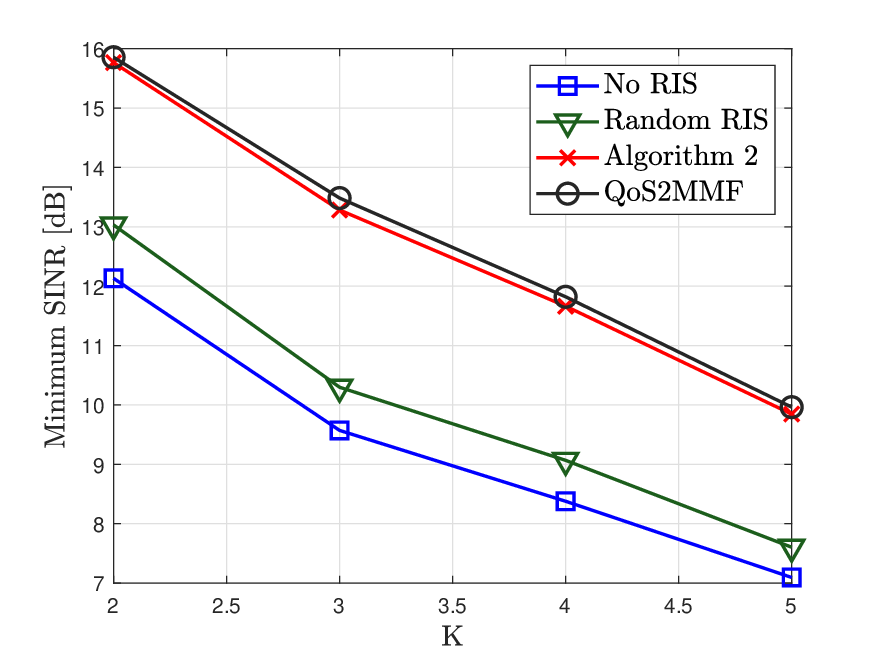}\\[-.5em]
        \centering
        \caption{\footnotesize Average minimum SINR  vs. $K$ ($N=32, M=100$, $G=2$).}
        \label{fig:MinSINRvsK}
         \vspace{-1em}
\end{figure}

 Fig.~\ref{fig:MinSINRvsN} shows the average minimum SINR vs. the number of BS antennas $N$. Since IP method is not effective, we exclude this method in comparison. Again, we see that    Algorithm \ref{alg2} provides about $3$ dB  SINR gain over random RIS under $M=100$ for different values of $N$, and its performance is close to that of QoS2MMF.
In Fig.~\ref{fig:MinSINRvsK}, we show the average minimum SINR as the number of users  $K$ per group increases. Again, Algorithm~\ref{alg2} and QoS2MMF perform similarly. The gain of Algorithm~\ref{alg2} over  no RIS or random RIS is consistent  for all values of $K$ with $2.5\sim 3$~dB gain. 

\section{Conclusion}\label{sec:conclusion}
In this paper, we consider joint BS and RIS beamforming design for  RIS-assisted downlink multicasting,  aiming to provide efficient and scalable algorithms. We show that the QoS problem can be  broken down into  a BS multicast beamforming QoS problem and an RIS passive multicast beamforming MMF problem, and we propose an AMBF approach to solve the two subproblems alternatingly. Our proposed fast algorithm for AMBF  utilizes the optimal BS mutlicast beamforming structure  for computational efficiency and employs a PSA-based fast first-order algorithm to solve the RIS passive multicast MMF problem via iterative closed-form updates. The overall computational complexity grows linearly with the number of BS antennas and RIS elements. For the RIS-assisted MMF problem, we utilize the optimal BS beamforming structure  to transform the MMF problem and propose an APSA  algorithm using all closed-form updates, leading to  a highly computationally efficient solution with complexity linear the number of BS antennas and RIS elements.
 Simulation results shows that  our proposed algorithms provide favorable performance with low cost in computational time as compared with  existing methods.

\appendices
\section{Proof of Proposition~\ref{prop1}}\label{appA}
\emph{Proof:}
We can equivalently rewrite $\Sc_o$ as
\begin{align*}
 \max_{\wbf}& \ \bigg(\max_{\ebf: |e_m|^2=1 ,  m \in \Mc}\min_{i, k} \frac{\SINR_{ik}}{\gamma_{ik}}\bigg) &
\st & \ \sum_{i=1}^G \|\wbf_i\|^2\le P.
\end{align*}
For the QoS problem  $\Pc_o$, consider its equivalent formulation  $\Pc_2$ in \eqref{P2} derived in Section~\ref{subsec:QoS}. Comparing the above problem and $\Pc_2$, we see that the objective functions and the constraints in these two problems have the same forms but switched, indicating that they  are the inverse problems (and equivalently $\Sc_o$ and $\Pc_o$ are the inverse problems). It is straightforward to see that, at the optimality, the respective constraints in $\Sc_o'$ and $\Pc_2$ are attained with equality. As a result, we have the relations in \eqref{inv_prob2}.
\endIEEEproof
\bibliographystyle{IEEEtran}
\bibliography{IEEEabrv,master,myBib}
\end{document}